\documentclass[aps,prl,reprint,groupedaddress]{revtex4-2}

\usepackage{graphicx} 
\usepackage{dcolumn}
\usepackage{bm}
\usepackage{commath} 
\usepackage{natbib}
\usepackage{xcolor}

\begin{document}

\title{Stretching Response of a Polymer Chain with Deformable Bonds}
\author{Jie Zhu}
\author{Laurence Brassart}
\affiliation{Department of Engineering Science, University of Oxford, Oxford OX1 3PJ, United Kingdom.}

\begin{abstract}
The stretching response of polymer chains fundamentally determines the mechanical properties of polymer networks. In this Letter, we develop a statistical mechanics model that incorporates both bond stretching and bond angle deformation, enabling accurate predictions of chain behavior up to large forces. We further propose a semianalytical deformable freely rotating chain (dFRC) model, which represents the chain as a freely rotating chain with effective bond stretch and bond angle that depend on the chain stretch. Using physical parameters without fitting, both the statistical and dFRC models achieve excellent agreement with experimental data for carbon chains across all force regimes. Additionally, the dFRC model provides a direct estimate of the bond force, which is important to predict chain scission. By capturing key bond deformations while remaining computationally efficient, our work lays the foundation for future modeling of polymer network elasticity and failure.
\end{abstract}

\maketitle

The molecular structure and conformations of polymer chains fundamentally determine the macroscopic mechanical properties of polymer networks. Accordingly, chain models serve as key building blocks for polymer network studies, including mean-field theories \cite{james1947statistical,storm2005nonlinear,feng1985effective}, discrete network models \cite{gusev2019numerical,alame2020effect,araujo2024micromechanical}, and continuum mechanics models  \cite{boyce2000constitutive,vernerey2017statistically,mulderrig2021affine}. When a polymer network is subjected to large deformations, such as during fracture, cavitation, or rapid loading, some chains can become highly stretched and eventually rupture. In these situations, the chain behavior is dominated by energetic contributions due to bond stretching and bond angle opening \cite{hugel2005highly,wang2013inherent,lu2022understanding}. Therefore, single chain models that are valid up to large stretches are needed. 

Classical entropic chain models, such as the freely jointed chain (FJC) and freely rotating chain (FRC) models \cite{rubinstein2003polymer}, do not account for energetic contributions and thus cannot capture the high-force behavior. On the other hand, the wormlike chain (WLC) model is commonly adopted to account for bending energy in semiflexible polymers \cite{marko1995stretching} but is limited to small equilibrium bond angles. Although extendable FJC and WLC models have been proposed \citep{balabaev2009extension,manca2012elasticity,fiasconaro2019analytical,buche2022freely,mulderrig2023statistical,mao2017rupture,fiasconaro2023elastic,netz2001strongly,kierfeld2004stretching,rosa2003elasticity}, they do not simultaneously accommodate both bond stretching and bond angle opening with a finite equilibrium bond angle, raising concerns about the physical meaning of their best-fit parameters and their relevance to chain scission. 

In this Letter, we address this limitation by developing a statistical mechanics model and a semianalytical model that explicitly incorporate both bond stretching and bond angle deformation in the stretching behavior of polymer chains. In the statistical model, we use the transfer-matrix (TM) technique to calculate the partition function in the Gibbs ensemble, from which we obtain the Gibbs free energy and force-extension relationship. Our results show that both bond stretching and bond angle opening significantly affect the chain stiffness at high forces. Using the TM calculations as reference, we further develop a semianalytical model by representing the polymer chain as a deformable FRC (dFRC) where the bond length and bond angle depend on the applied chain stretch, inspiring from original ideas in Refs \cite{mao2017rupture,lavoie2019modeling}. This is achieved by developing a new explicit formula for the classical FRC, which is shown to be valid across the entire force range. Using molecular mechanics-derived parameters without any fitting, both the statistical and dFRC models accurately predict the force-extension behavior of carbon chains and exhibit excellent agreement with experimental data over all force regimes. By capturing key bond-level deformations while maintaining computational efficiency, the dFRC model provides a foundation for advancing polymer network modeling, particularly in scenarios involving failure by chain scission, where bond force predictions are important.

\begin{figure}
    \includegraphics{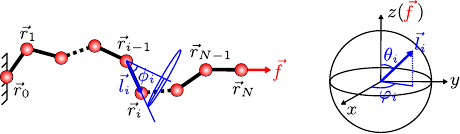}
    \caption{A polymer chain with $N$ bonds subjected to a constant force $\vec{f}$ at $\vec{r}_N$. The other chain end at $\vec{r}_0$ is fixed. The orientation of each bond is described using spherical coordinates, with the $z$ axis aligned with the force.}
    \label{fig:schematicofchain}
\end{figure}

We consider a polymer chain composed of $N$ bonds (Fig.~\ref{fig:schematicofchain}). The conformation of the chain is described by the positions $\vec{r}_i$ of each atom $(i=0,\ldots,N)$. The chain is subjected to a prescribed force $\vec{f}$ at the terminal atom $\vec r_N$, with the other terminal atom fixed at the origin $\vec{r}_0=(0,0,0)$. Each bond vector, $\vec{l}_i=\vec{r}_i-\vec{r}_{i-1}$, $(i=1,\ldots,N)$, is represented in spherical coordinates as $\vec{l}_i=l_i\left(\sin{\theta_i}\cos{\varphi_i},~\sin{\theta_i}\sin{\varphi_i},~\cos{\theta_i}\right)$, where $l_i$ is the bond length, $\theta_i$ is the polar angle relative to the direction of the applied force $\vec{f}$, and $\varphi_i$ is the azimuthal angle. The bond angle between adjacent bonds, $\phi_i=\angle(\vec{l}_{i-1},\vec{l}_i)$, ($i=2,\ldots,N$), is calculated as $\phi_i = \arccos{\left[\sin{\theta_i}\sin{\theta_{i-1}}\cos{\omega_i}+\cos{\theta_i}\cos{\theta_{i-1}}\right]}$, where $\omega_i = \varphi_i-\varphi_{i-1}$ is the difference in azimuthal angle between two adjacent bonds. 

The partition function for the chain in the Gibbs ensemble is given by (Supplemental Material, Sec. S1 \cite{supplemental}\nocite{talamini2018progressive,li2020variational,arunachala2023multiscale,arunachala2024multiscale}):
\begin{equation}
     Z \propto \int e^{-\frac{1}{k_BT}\left[\sum_{i=1}^{N}v_{\text{str}}{\left(l_{i}\right)} + \sum_{i=2}^{N}v_{\text{ben}}{\left(\phi_i\right)} - \sum_{i=1}^{N}fl_i\cos{\theta_i}\right]}\,dq,
     \label{eq:Zf}
\end{equation}
where $q=(\vec{l}_1,\ldots,\vec{l}_{N})$, $k_B$ is the Boltzmann constant, $T$ is the temperature, and $f = \norm{\vec{f}}$. The bond stretching energy is assumed quadratic for simplicity: $v_{\text{str}}{\left(l_i\right)} = \frac{1}{2}k_{l}\left(l_i-l_e\right)^2$, where $l_e$ is the equilibrium bond length, and $k_{l}$ is the bond stretching stiffness. Similarly, the bond angle deformation energy (``bending energy") is taken as $v_{\text{ben}}{\left(\phi_i\right)} = \frac{1}{2}k_{\phi}\left(\phi_i-\phi_e\right)^2$, where $\phi_e$ is the equilibrium bond angle, and $k_\phi$ is the bond bending stiffness. We neglect the bond rotation energy, as it is typically much smaller than the energies associated with bond stretching and bending in homogeneous backbone chains (e.g., carbon-carbon backbone) \cite{wang2013inherent}, so that bonds can freely rotate. In addition, we neglect the long-range interactions between atoms that are far apart along the chain contour, so that the chain is an ideal chain \cite{rubinstein2003polymer}.

The bending energy couples the configurations of two consecutive bonds. Consequently, the Gibbs partition function [Eq. (\ref{eq:Zf})] cannot be factorized into $N$ identical terms, different from the extensible FJC model. To address this difficulty, we employ the numerical TM method \cite{baxter2016exactly,ross2014introduction}, building on the work of Livadaru \textit{et al}., who applied this technique to the FRC model \cite{livadaru2003stretching}. The weighted probability density of finding the ($i+1$)th bond at an angle $\theta$ can be expressed as (Supplemental Material, Sec. S2.1 \cite{supplemental}):
\begin{equation}
    W_{i+1}(\theta) = \int {P_i\left(\theta'\right)\textrm{T}\left(\theta,\theta'\right)\,d\theta'},
    \label{eq:Wi+1}
\end{equation}
where $P_i(\theta) = W_{i}\left(\theta\right)/[\int W_{i}(\theta')\,d\theta'] $ is the bond orientation probability density of the $i$th bond, and $\int P_i(\theta)d\theta = 1$. The transfer operator $\textrm{T}\left(\theta,\theta'\right)$ characterizes the interaction between the $(i+1)$th bond at angle $\theta$ and the $i$th bond at angle $\theta'$, which is given by
\begin{equation}
    \textrm{T}\left(\theta,\theta'\right) = \iint e^{-\frac{1}{k_BT}\left[v_{\text{str}}{\left(l\right)} + v_{\text{ben}}{\left(\phi\right)} - fl\cos{\theta}\right]}l^2\sin{\theta}\,d\omega\,dl,
\end{equation}
where $l$ is the length of the $(i+1)$th bond, $\phi$ is the bond angle between the $(i+1)$th and $i$th bonds, and $\omega$ is the difference in their azimuthal angles. The weighted probability density of the first bond is given by
\begin{equation}
    W_1(\theta) = \iint e^{-\frac{1}{k_BT}\left[v_{\text{str}}{(l)} - fl\cos{\theta}\right]}l^2\sin{\theta}\,d\omega\,dl
    \label{eq:W1} 
\end{equation}
where $\omega$ is the azimuthal angle of the first bond. Starting with $W_1(\theta)$, the weighted probability density of each subsequent bond can be determined iteratively. Finally, the Gibbs partition function [Eq. (\ref{eq:Zf})] can be obtained as (Supplemental Material, Sec. S2.1 \cite{supplemental})
\begin{equation}
    \label{eq:Zf2}
     Z \propto \prod_{i=1}^N\left[\int W_i(\theta)\,d\theta\right].
\end{equation}
The chain end-to-end distance $r$ is derived from the Gibbs partition function as
\begin{equation}
     r = k_BT\frac{\partial}{\partial f}\ln{Z} = -\frac{\partial G}{\partial f},
\end{equation}
where $G = -k_BT\ln{Z}$ is the Gibbs free energy. Note that $r=\norm{\Vec{r}}=\norm{\left<\vec{r}_N\right>}$ is the norm of the ensemble average of the chain end-to-end vector. The partition function is calculated using quadrature methods, and the force-extension relationship is obtained by numerical differentiation (Supplemental Material, Sec. S2.1 \cite{supplemental}). The special cases of freely jointed bonds ($k_\phi=0$) and fixed bond angles ($k_\phi\rightarrow\infty$) are discussed in Supplemental Material, Secs. S2.2 and S2.3, respectively. Bond orientation probability densities $P_i(\theta)$ are illustrated in Sec. S3, where we show that $P_i(\theta)$ converges to a stable distribution $P_\infty(\theta)$ as the bond index $i$ increases. For sufficiently large $N$, the Gibbs partition function [Eq. (\ref{eq:Zf2})] can be approximated as $Z \propto \left[\int W_{\infty}(\theta)\,d\theta\right]^N$, where $W_{\infty}(\theta) = \int {P_{\infty}\left(\theta'\right)\textrm{T}\left(\theta,\theta'\right)\,d\theta'}$ is the converged weighted probability density. The resulting force-extension relationship becomes independent of $N$, corresponding to the thermodynamic limit. The effect of $N$ on the force-extension relationship is further discussed in Supplemental Material, Sec. S4.1.

We examine how bond stretching and bond angle opening influence the stretching response in the thermodynamic limit ($N \rightarrow \infty$). To isolate their individual effects, we analyze the following limiting cases:

\begin{figure}
    \includegraphics{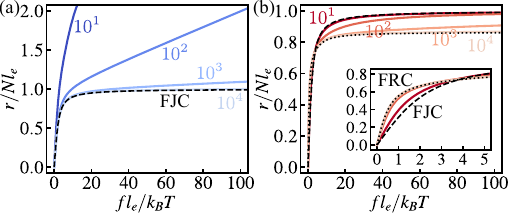}
    \caption{Effect of bond deformation on the stretching response of polymer chains. (a) Chains with freely jointed extensible bonds ($k_\phi=0$; $k_ll_e^2/k_BT=10^1$, $10^2$, $10^3$, $10^4$). (b) Chains with rigid bonds ($k_l l^2_e/k_B T\rightarrow\infty$) and deformable bond angles ($\phi_e=60^{\circ}$; $k_\phi\pi^2/k_BT=10^1$, $10^2$, $10^3$, $10^4$).}
    \label{fig:StatisticalModel}
\end{figure}

(i) Freely jointed extensible bonds. Force-extension curves are shown in Fig.~\ref{fig:StatisticalModel}(a) for different values of the normalized bond stretching stiffness, $k_l l_e^2/k_B T$. At low forces, the chain stretching response is primarily governed by entropic elasticity and follows the Gaussian chain model $r=Nl_e^2f/(3k_BT)$, where $Nl_e^2/(3k_BT)$ represents the entropic spring constant \cite{rubinstein2003polymer}. As the force increases, bond stretching results in a second elastic regime with chain stiffness proportional to $k_l$. In the limit $k_l l^2_e/k_B T \rightarrow\infty$, the extensible FJC model recovers the FJC model with contour length $R_{\text{max}}=Nl_e$. These results are consistent with the findings in Ref. \cite{manca2012elasticity}.

(ii) Rigid bonds and deformable bond angles. Force-extension curves are shown in Fig.~\ref{fig:StatisticalModel}(b) for $\phi_e=60^{\circ}$ and different values of the normalized bond bending stiffness, $k_{\phi} \pi^2 /k_B T$. For $k_\phi=0$, the model recovers the FJC model. For $k_\phi \pi^2/k_B T \rightarrow\infty$, the model tends to the FRC model with contour length $R_{\text{max}}=Nl_e\cos(\phi_e/2)$. At low applied forces, increasing $k_\phi$ softens the chain since the coiled configuration becomes less probable. As the force increases, a larger $k_\phi$ stiffens the chain as it approaches its extensibility limit. This stiffening effect depends on the equilibrium bond angle $\phi_e$, with a larger $\phi_e$ leading to a reduced contour length, see Supplemental Material, Sec. S4.2 \cite{supplemental}. 

While the statistical model effectively captures the effect of bond deformation on the stretching response of polymer chains, its lack of a closed-form expression limits practical use in single-molecule experiments and polymer network modeling. To address this, we further develop a semianalytical approximation. Following the approach proposed by Lavoie \textit{et al}. \cite{lavoie2019modeling}, we assume uniform and nonfluctuating bond length $l$ and bond angle $\phi$. The Helmholtz free energy $\Psi$ of a polymer chain under a prescribed end-to-end distance $r$ is then given by   
\begin{equation}\label{eq:energy_simplified}
    \Psi = Nv_{\text{str}}\left(l\right) + (N-1)v_{\text{ben}}\left(\phi\right) + \Psi_{\text{ent}}\left(r,l,\phi\right),
\end{equation}
where the first two terms, respectively, represent the energetic contributions due to bond stretching and bond angle deformation, and the third term $\Psi_{\text{ent}}$ is the entropic contribution. In Ref. \cite{lavoie2019modeling}, $\Psi_{\text{ent}}$ was obtained by integrating the closed-form force-extension relation of the WLC model, where the persistence length was expressed in terms of $l$ and $\phi$. However, this WLC representation introduces additional approximations which are difficult to quantify. Here, in contrast, we directly take $\Psi_{\text{ent}}$ as the free energy of a FRC with bond length $l$ and bond angle $\phi$, consistent with the decomposition in Eq. (\ref{eq:energy_simplified}). The bond parameters $l$ and $\phi$ are determined by minimizing the Helmholtz free energy at a prescribed chain end-to-end distance $r$, which implies $\left({\partial \Psi}/{\partial l}\right)_{r,\phi} = \left({\partial \Psi}/{\partial \phi}\right)_{r,l} = 0$. The force-displacement response is then derived as $f = {d\Psi}/{dr} = \left({\partial \Psi_{\text{ent}}}/{\partial r}\right)_{l,\phi}$. We refer to this model as the ``deformable FRC" (dFRC). 

The practical implementation of the dFRC model requires an explicit expression for the force-extension relationship of the FRC, valid across the entire force range. However, existing formulations are either limited to specific force ranges or implicit  \cite{livadaru2003stretching,hugel2005highly}. To address this, we propose a new explicit formula for the FRC, guided by its limiting behaviors at small and large deformations (Supplemental Material, Sec. S5 \cite{supplemental}):
\begin{equation}
    f = \frac{k_BT}{l_k}{\left\{\beta + \frac{1}{2}\frac{\left(r^*\right)^2}{\left(1-r^*\right)^2}\left[1-\left(r^*\right)^{l_k/l-1}\right]\right\}},
    \label{eq:simplified_model}
\end{equation}
where $\beta=\mathcal{L}^{-1}(r^*)$ and $\mathcal{L}(x)=\coth{(x)-1/x}$ is the Langevin function. Here, $r^*=r/R_{\text{max}}$ represents the relative end-to-end distance, with the contour length given by $R_{\text{max}} = Nl\cos{\left(\phi/2\right)}$. The equivalent Kuhn length is $l_k =\left<r^2\right>/R_{\text{max}}= {2l\cos{\left(\phi/2\right)}}/(1-\cos{\phi})$, where the mean-square end-to-end distance is $\left<r^2\right>=Nl^2(1+\cos{\phi})/(1-\cos{\phi})$ \cite{rubinstein2003polymer}. Equation (\ref{eq:simplified_model}) shows a remarkable agreement with TM calculations across a wide range of bond angles ($\phi=10^\circ\text{--}90^\circ$), maintaining a relative error below $6\%$ over the entire force range ($fl_e/k_BT=10^{-3.5}\text{--}10^{2.5}$) (Supplemental Material, Fig. S10 \cite{supplemental}). The FRC free energy $\Psi_{\text{ent}}(r,l,\phi)$ is obtained by integrating the force-extension relation as $\Psi_{\text{ent}}(r,l,\phi) = \int{f}dr$, giving
\begin{eqnarray}\label{eq:Psi_ent}
    \Psi_{\text{ent}} = && \,k_BT\frac{R_{\text{max}}}{l_k}\left[r^*\beta + \ln{\frac{\beta}{\sinh{\beta}}}+\frac{1+r^*(1-r^*)}{2(1 - r^*)}\right.\nonumber\\
    &&\left. +\ln(1 - r^*)- \frac{1}{2}\mathcal{B}\left(r^*; 2 + l_k/l, -1\right)\right],
\end{eqnarray}
where $\mathcal{B}\left(x;a,b\right)=\int_0^x{t^{a-1}\left(1-t\right)^{b-1}\,dt}$ is the incomplete beta function. For freely jointed bonds ($k_\phi=0$), only bond stretching is considered, with the bond length $l$ determined by energy minimization. In this case, the Kuhn length and contour length are $l_k=l$ and $R_{\text{max}}=Nl$, respectively. The force-extension relationship [Eq. (\ref{eq:simplified_model})] recovers the FJC model with $f = ({k_BT}/{l_k})\beta$, and the entropic energy [Eq. (\ref{eq:Psi_ent})] becomes $\Psi_{\text{ent}}(r,l)=Nk_BT\left[r^*\beta + \ln{\frac{\beta}{\sinh{\beta}}}\right]$. This recovers the semianalytical extensible FJC model in Ref. \cite{mao2017rupture}. 

\begin{figure}
    \includegraphics{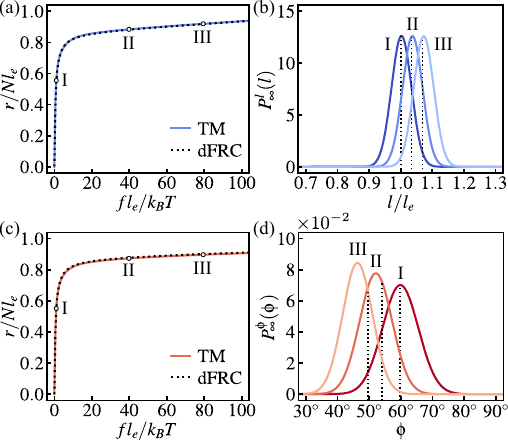}
    \caption{Comparison of statistical (TM) and dFRC models for chains with deformable bonds. (a) Force-extension curves for a chain with extensible bonds ($k_ll_e^2/k_BT=10^3$) and fixed bond angles ($k_\phi \pi^2/k_B T \rightarrow\infty$, $\phi_e=60^\circ$). (b) Converged bond length distribution in the statistical model at different forces from (a). (c) Force-extension curves for a chain with rigid bonds ($k_l l^2_e/k_B T\rightarrow\infty$) and deformable bond angles ($k_\phi\pi^2/k_BT=10^3$, $\phi_e=60^\circ$). (d) Corresponding converged bond angle distribution from (c). In (b) and (d), black dotted lines indicate the optimized bond lengths or angles in the dFRC model.} 
    \label{fig:ComparisonStatisticalAndSimplified}
\end{figure}

We validate the semianalytical dFRC model for a polymer chain with extensible bonds and deformable bond angles by comparing its predicted force-extension relationships to reference predictions from the statistical model (TM calculations). Additionally, we assess the model core assumptions of uniform and nonfluctuating bond length and bond angle by analyzing corresponding probability densities obtained from TM calculations (Supplemental Material, Sec. S6 \cite{supplemental}). The semianalytical extensible FJC model is similarly validated against the statistical model in Supplemental Material, Sec. S6.1. Here, we consider two representative cases:

(i) Extensible bonds and fixed bond angles. In the statistical model, the bond length probability density $P_i^l(l)$ depends on the bond index $i$ but quickly converges to $P_\infty^l(l)$ as $i$ increases (Supplemental Material, Fig. S13 \cite{supplemental}), supporting the uniform bond length assumption. This converged distribution depends on the bond stretching stiffness $k_l$, which controls the length variance, and the applied force $f$, which shifts the average bond length (Supplemental Material, Fig. S14 \cite{supplemental}). As shown in Fig. \ref{fig:ComparisonStatisticalAndSimplified}(b), for $k_ll_e^2/k_BT=10^3$, $P_\infty^l(l)$ remains concentrated around its most probable value, supporting the nonfluctuating bond length assumption. The optimized bond length in the dFRC model closely follows the peak of $P_\infty^l(l)$, yielding an accurate force-extension prediction [Fig. \ref{fig:ComparisonStatisticalAndSimplified}(a)]. Even for $k_ll_e^2/k_BT=10^1$, where the bond length is more widely distributed, the dFRC model still approximates the force-extension response well (Supplemental Material, Fig. S15 \cite{supplemental}).

(ii) Rigid bonds and deformable bond angles. With increasing bond index $i$, the bond angle probability density $P_i^\phi(\phi)$ given by the statistical model quickly converges to $P_\infty^\phi(\phi)$ (Supplemental Material, Fig. S17 \cite{supplemental}), supporting the uniform bond angle assumption. This converged distribution depends on the bond bending stiffness $k_\phi$ and the applied force $f$ (Supplemental Material, Fig. S18 \cite{supplemental}). For $k_\phi\pi^2/k_BT=10^3$, $P_\infty^\phi(\phi)$ is narrowly distributed, supporting the nonfluctuating bond angle assumption [Fig. \ref{fig:ComparisonStatisticalAndSimplified}(d)]. As $f$ increases, $P_\infty^\phi(\phi)$ narrows further and shifts to the left, indicating a reduction in both the average and variability of the bond angles. At lower forces, the optimized bond angle in the dFRC model aligns closely with the peak of $P_\infty^\phi(\phi)$, corresponding to the most probable bond angle. However, at higher forces, the optimized value overestimates the most probable value [Fig. \ref{fig:ComparisonStatisticalAndSimplified}(d)]. Nevertheless, the dFRC model still accurately predicts the force-extension relation [Fig. \ref{fig:ComparisonStatisticalAndSimplified}(c)]. The accuracy of the dFRC model also depends on $k_\phi$. As $k_\phi$ decreases, the bond angle distribution broadens, and the dFRC model performance deteriorates, becoming inapplicable for $k_\phi\pi^2/k_BT=10$ (Supplemental Material, Fig. S19 \cite{supplemental}).

\begin{figure}
    \includegraphics{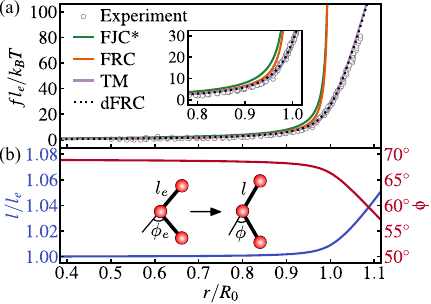}
    \caption{\label{fig:CarbonChains}Stretching response of carbon chains. (a) Comparison of predictions from the FJC*, FRC, statistical (TM) and dFRC models against experimental data from \cite{wang2013inherent}. (b) Evolution of bond length $l$ and bond angle $\phi$ in the dFRC model. $R_0$ denotes the contour length at zero force.}
\end{figure}

Finally, we apply our statistical and dFRC models to polymer chains with a carbon-carbon backbone, such as polyethylene, polystyrene, and polypropylene. Molecular mechanics calculations indicate that these chains have an equilibrium bond length of $l_e=1.53~\text{\AA}$ and an equilibrium bond angle of $\phi_e=69^{\circ}$, with bond stiffness $k_l=4.29\times10^{-18}~\text{J}/\text{\AA}^2$ for stretching and $k_\phi=7.47\times10^{-19}~\text{J}/\text{rad}^2$ for bending \cite{sorensen1988prediction}. Using these parameters, we simulate chain stretching at room temperature ($T=23^\circ$) and compare predictions with experimental data from \cite{wang2013inherent}. For reference, we also include the equivalent FJC model (FJC*), which represents the chain using rigid Kuhn segments of length $l_k={2l_e\cos{\left(\phi_e/2\right)}}/(1-\cos{\phi_e})$, and the FRC model, which considers rigid bonds of length $l_e$ and fixed bond angles at $\phi_e$. As shown in Fig. \ref{fig:CarbonChains}(a), all models agree well with experimental data at small deformations ($r/R_0\leq0.8$), where the response is entropy dominated. Here, $R_0=Nl_e\cos{(\phi_e/2)}$ is the contour length at zero force. At intermediate forces ($0.8<r/R_0\leq0.9$), the FRC model outperforms the FJC* model by incorporating fixed bond angles, thereby better capturing the effect of angular constraints on chain conformations. However, at large deformations ($r/R_0>0.9$), the FRC model diverges from experimental data as bond deformations become significant. Additional comparisons with extensible FJC and FRC models reveal that using physical parameters alone fails to capture the large deformation response (Supplemental Material, Sec. S7 \cite{supplemental}). This highlights the necessity of incorporating both bond stretching and bond angle opening in chain elasticity models. In contrast, our statistical (TM) and dFRC models remain accurate across the entire force range by capturing these energetic effects [Fig. \ref{fig:CarbonChains}(a)]. Both bond stretching and bond angle opening contribute significantly to the response, as indicated by the dimensionless bond parameters $k_ll_e^2/k_BT=2466$ and $k_\phi\pi^2/k_BT=1810$, which strongly influence chain stiffness at high forces (Fig. \ref{fig:StatisticalModel}). These parameters also ensure the close agreement between the dFRC and statistical models (Fig. \ref{fig:ComparisonStatisticalAndSimplified}). 

Beyond improving chain-level force-extension predictions, the dFRC model also provides access to internal bond-level variables. As shown in Fig. \ref{fig:CarbonChains}(b), it predicts the evolution of bond length and bond angle with increasing chain extension. For extension ratios $r/R_0\leq0.9$, the response is entropy dominated, with bond lengths and angles close to their equilibrium values. Consequently, the force-extension curve in this regime aligns with the FRC model [Fig. \ref{fig:CarbonChains}(a)]. At larger extensions, energetic contributions become significant, leading to bond stretching and bond angle opening. Based on the predicted bond length, the bond force can be obtained as $f_b=\partial{v_{\text{str}}(l)}/\partial{l}=k_l(l-l_e)$. In the entropy-dominated regime, $f_b$ remains nearly zero due to negligible bond stretching, but it increases as deformation progresses, potentially leading to bond breakage and chain scission \cite{grandbois1999strong,ghatak2000interfacial,chaudhury1999rate,lavoie2019modeling,guo2021micromechanics}.

In summary, we have presented two complementary models that accurately capture the stretching behavior of polymer chains up to large deformations by incorporating both bond stretching and bond angle opening simultaneously. Without relying on parameter fitting, both models show excellent agreement with experimental data on carbon chain stretching across all force regimes. The semianalytical dFRC model, in particular, balances accuracy with computational efficiency and provides bond-level resolution, making it suitable for integration into polymer network simulations aimed at predicting elasticity and failure. While we adopted quadratic forms for bond stretching and bending energies, our framework is not restricted to these choices and can accommodate more general potentials, such as Lennard-Jones or Morse potentials. It can also be extended to hybrid chains with multiple backbone bond types or less flexible chains with large side groups and strong rotation hindrance. Overall, this work provides a robust foundation for advancing physically grounded and predictive modeling of polymer network mechanics, particularly under extreme deformation and failure conditions.

\textit{Acknowledgments}---The authors thank Lucas Mangas Araujo from University of Oxford and Kangjie Zheng from Peking University for important discussions and helpful comments on the manuscript. J.Z. acknowledges the support of an EPSRC DTP studentship at the University of Oxford. L.B. acknowledges supports from UKRI through a Future Leaders Fellowship [MR/W006995/1].

\textit{Data availability}---The data that support the findings of this Letter are openly available \cite{Zhu617}.

\bibliography{bibfile}

\end{document}


\title{Supplementary Material: Stretching Response of a Polymer Chain with Deformable Bonds}
\author{Jie Zhu}
\author{Laurence Brassart}
\affiliation{Department of Engineering Science, University of Oxford, Oxford OX1 3PJ, United Kingdom.}

\maketitle

\renewcommand{\theequation}{S\arabic{equation}}
\renewcommand{\thefigure}{S\arabic{figure}}
\renewcommand{\thetable}{S\arabic{table}}
\renewcommand{\bibnumfmt}[1]{[S#1]}
\renewcommand{\citenumfont}[1]{S#1}

\section{S1 Theoretical Framework of the Statistical Model}
Consider a polymer chain consisting of $N+1$ atoms connected by $N$ bonds (Fig. 1). In the Gibbs ensemble, the first atom is fixed at position $\vec{r}_0=(0,0,0)$, and the last atom is subjected to a prescribed force $\vec{f}$. The chain configuration is described by the atom positions $\vec{r}_i\left(i=1,\ldots,N\right)$ and momenta $\vec{p}_i\left(i=1,\ldots,N\right)$. The dynamics of the system is described by the augmented Hamiltonian:
\begin{equation}
    \tilde{h}\left(\vec{r}_1,\ldots,\vec{r}_{N}, \vec{p}_1,\ldots,\vec{p}_{N},\vec{f}\right) = T\left(\vec{p}_1,\ldots,\vec{p}_{N}\right) + V\left(\vec{r}_1,\ldots,\vec{r}_{N}\right) - \vec{f}\cdot\vec{r}_{N}
\end{equation}
where $T\left(\vec{p}_1,\ldots,\vec{p}_{N}\right) = \sum_{i=1}^{N}[{\vec{p}_i\cdot\vec{p}_i}/{(2m)}]$ is the kinetic energy with $m$ the mass of an atom, $V\left(\vec{r}_1,\ldots,\vec{r}_{N}\right)$ is the interaction energy of the atoms, and $-\vec{f}\cdot\vec{r}_{N}$ is the potential energy of the applied force.

In thermal equilibrium with a reservoir at temperature $T$, the chain properties are described by the canonical ensemble distribution:
\begin{equation}
     \rho\left(q,p;\vec{f},T\right) = \frac{1}{Z}e^{-\frac{\tilde{h}\left(q,p,\vec{f}\right)}{k_BT}}
\end{equation}
where $q\equiv \left(\vec{r}_1,\ldots,\vec{r}_{N}\right)$, $p \equiv \left(\vec{p}_1,\ldots,\vec{p}_{N}\right)$, $k_B$ is the Boltzmann constant, and $Z$ is the canonical partition function in the Gibbs ensemble:
\begin{equation}
     Z\left(\vec{f},T\right) = \iint {e^{-\frac{\tilde{h}\left(q,p,\vec{f}\right)}{k_BT}}}\,dq\,dp = \left(2\pi mk_BT\right)^{{3N}/2}\int e^{-\frac{1}{k_BT}\left[V\left(q\right)-\vec{f}\cdot\vec{r}_{N}\right]}\,dq
\end{equation}
where we have integrated over the momenta. Since ${\partial\tilde{h}}/{\partial\Vec{f}} =-\vec{r}_N$, the mean position of the last atom is given by $\Vec{r}=\left<\vec{r}_N\right>=-\left<{\partial\tilde{h}}/{\partial\Vec{f}}\right>$, where $\left< \cdot \right> \equiv \iint \cdot \,dq\,dp$. Alternatively, the force-extension relation can directly be obtained from the partition function as \cite{manca2012elasticity}:
\begin{equation}
     \Vec{r}\left(\Vec{f},T\right) = k_BT\frac{\partial }{\partial \Vec{f}}\ln{Z} = -\frac{\partial G\left(\vec{f},T\right)}{\partial \Vec{f}} 
\end{equation}
where $ G\left(\vec{f},T\right)=-k_BT\ln{Z}$ is the Gibbs free energy of the chain.

For further analysis, we introduce the bond vectors $\vec{l}_i=\vec{r}_i-\vec{r}_{i-1}$, $(i=1,\ldots,N)$ and represent them in spherical coordinates, $\vec{l}_i=l_i\left(\sin{\theta_i}\cos{\varphi_i},~\sin{\theta_i}\sin{\varphi_i},~\cos{\theta_i}\right)$, as shown in Fig. 1. Here, $l_i$ is the length of the $i$th bond, $\theta_i\in[0,\pi]$ is the polar angle with respect to the direction of applied force $\vec{f}$, and $\varphi_i\in[0,2\pi)$ is the azimuthal angle. The bond angle between the $i$th and $(i-1)$th bonds can be expressed as:
\begin{equation}
    \phi_i = \angle(\vec{l}_{i-1},\vec{l}_i) =  \arccos{\left[\sin{\theta_i}\sin{\theta_{i-1}}\cos{\omega_i}+\cos{\theta_i}\cos{\theta_{i-1}}\right]}
\end{equation}
where $\omega_i = \varphi_i-\varphi_{i-1}$ is the difference in azimuthal angle between these two bonds. The interaction potential of the atoms is taken as:
\begin{equation}
    V = \sum_{i=1}^{N}v_{str}{\left(l_i\right)} + \sum_{i=2}^{N}v_{ben}{\left(\phi_i\right)}
\end{equation}
where $v_{str}{\left(l\right)}$ is the energy for stretching the bond between two adjacent atoms, and $v_{ben}(\phi)$ is the energy for deforming the angle between two consecutive bonds. In this work, we adopt the simple quadratic expressions: $v_{str}(l) = \frac{1}{2}k_{l}\left(l-l_e\right)^2$ and $v_{ben}{\left( \phi \right)} = \frac{1}{2}k_{\phi}\left(\phi-\phi_e\right)^2$, where $l_e$ and $\phi_e$ are the equilibrium bond length and bond angle, respectively. $k_l$ and $k_\phi$ are the stiffness for bond extension and bond angle deformation, respectively. Throughout this work, we do not consider the torsional energy of the bonds, i.e. we assume that bonds can freely rotate. The partition function becomes:
\begin{equation}
     Z \left(f,T\right) = \left(2\pi mk_BT\right)^{{3N}/2}\int e^{-\frac{1}{k_BT}\left[\sum_{i=1}^{N}v_{str}{\left(l_{i}\right)} + \sum_{i=2}^{N}v_{ben}{\left(\phi_i\right)} - \sum_{i=1}^{N} f l_{i} \cos \theta_i \right]}\,dq'
     \label{Seq:Zf}
\end{equation}
where $q'\equiv (\vec{l}_1,\ldots,\vec{l}_{N})$ and $f = \norm{\vec{f}}$. 

\section{S2 Transfer-Matrix Technique}
\subsection{S2.1 General solution}
\begin{figure}
    \includegraphics{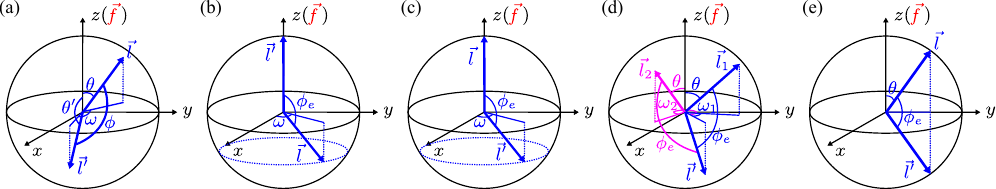}
    \caption{\label{sfig:Schematic_of_bond_angle}Spherical coordinates used in the transfer-matrix schemes for (a) a deformable bond angle $\phi$ and (b-e) a fixed bond angle $\phi_e$. (b) $\theta'=0$, $\theta=\phi_e$, and $\omega$ can be any value in the range of $[0,2\pi)$. (c) $\theta=0$, $\theta'=\phi_e$, and $\omega$ can be any value in the range of $[0,2\pi)$. (d) $\abs{\cos{\omega}}<1$, and $\omega$ has two possible values. (e) $\omega=0$, and $\theta'=\theta+\phi_e$.}
\end{figure}

Direct calculation of the partition function (Eq. \ref{Seq:Zf}) is challenging due to the dependence of the interaction potential on the bond angle $\phi$, which introduces a coupling between the orientations of two consecutive bonds. To address this difficulty, we adopt the transfer-matrix (TM) technique \cite{baxter2016exactly}, inspiring from Ref. \cite{livadaru2003stretching}. The partition function is rewritten in the equivalent form:
\begin{equation}
    Z \left(f,T\right) = \left(2\pi mk_BT\right)^{{3N}/2}\int B_N(\theta)\,d\theta
    \label{Seq:Zf2}
\end{equation}
where the function $B_i(\theta)$, ($i=1,\ldots,N$) measures the probability of finding the $i$th bond at a certain angle $\theta$ with respect to the direction of the applied force. It is obtained via the recursive formula:
\begin{equation}
    B_{i+1}(\theta) = \int B_i(\theta')\textrm{T}\left(\theta,\theta'\right)\,d\theta'
    \label{Seq:Bi+1}
\end{equation}
with the initial distribution
\begin{equation}
    B_{1}(\theta) = 2\pi \int e^{-\frac{1}{k_BT}\left[v_{str}{\left(l\right)} - fl\cos{\theta}\right]} l^2\sin{\theta}\,dl
    \label{Seq:B1}
\end{equation}
$\textrm{T}(\theta,\theta')$ is the transfer operator describing the interaction between two consecutive bonds at angles $\theta$ and $\theta'$. It is given by:
\begin{equation}
    \textrm{T}\left(\theta,\theta'\right) = \iint e^{-\frac{1}{k_BT}\left[v_{str}{\left(l\right)} + v_{ben}{\left(\phi\right)} - fl\cos{\theta}\right]} l^2\sin{\theta}\,d\omega\,dl
    \label{Seq:T}
\end{equation}
where $\phi=\arccos{[\sin\theta\sin\theta'\cos\omega+\cos\theta\cos\theta']}$ and $\omega$ is the difference in azimuthal angle between these two bonds (Fig. \ref{sfig:Schematic_of_bond_angle}a). Mathematically, Eq. \ref{Seq:Bi+1} resembles a Chapman-Kolmogorov equation, with $\textrm{T}\left(\theta,\theta'\right)$ serving as a conditional probability density \cite{ross2014introduction}. Starting from $B_1(\theta)$, the $B_i(\theta)$ of each subsequent bonds can be computed iteratively, and the partition function can be obtained from Eq. \ref{Seq:Zf2}.

To keep the high-dimensional integration within a numerically manageable range, it is convenient to define the rescaled function:
\begin{equation}\label{Seq:rescale}
    W_{i+1}(\theta) = \frac{B_{i+1}(\theta)}{\int{B_i(\theta')\,d\theta'}}
\end{equation}
For the first bond, $W_{1}(\theta) = B_{1}(\theta)$. Substituting Eq. \ref{Seq:Bi+1} into Eq. \ref{Seq:rescale} gives:
\begin{equation}
    W_{i+1}(\theta) = \int {P_i\left(\theta'\right)\textrm{T}\left(\theta,\theta'\right)\,d\theta'}
    \label{Seq:Wi+1}
\end{equation}
where $P_i(\theta)$ is the probability density of finding the $i$th bond at an angle $\theta$:
\begin{equation}\label{Seq:Pi}
    P_i(\theta) = \frac{B_{i}(\theta)}{\int B_{i}(\theta')\,d\theta'} = \frac{W_{i}(\theta)}{\int W_{i}(\theta')\,d\theta'}
\end{equation}
It is clear that $\int P_i(\theta)\, d\theta =1$. The original probability function $B_{i+1}(\theta)$ can then be expressed as:
\begin{equation}
    B_{i+1}(\theta) = W_{i+1}(\theta)\prod_{j=1}^i{\left[\int W_j(\theta)\,d\theta\right]}
\end{equation}
and the partition function (Eq. \ref{Seq:Zf2}) becomes:
\begin{equation}
     Z\left(f,T\right) = \left(2\pi mk_BT\right)^{{3N}/2}\prod_{i=1}^N\left[\int W_i(\theta)\,d\theta\right]
\end{equation}
Finally, the Gibbs free energy can be expressed as:
\begin{equation}
     G\left(f,T\right) = -k_BT\sum_{i=1}^N\ln{\left[\int W_i(\theta)\,d\theta\right]} 
     \label{Seq:G}
\end{equation}
up to an additive constant independent of $f$. The force-extension relation of the chain is given as:
\begin{equation}
     r\left(f,T\right) = -\frac{\partial G\left(f,T\right)}{\partial f}
     \label{Seq:rf}
\end{equation}
where $r=\norm{\vec{r}}$ is the chain end-to-end distance.

In practice, the functions $W_{i+1}(\theta)$ are calculated iteratively via the probabilities $P_i(\theta)$ using Eqs \ref{Seq:Wi+1} and \ref{Seq:Pi}, starting from $W_{1}(\theta)=B_1(\theta)$, Eq. \ref{Seq:B1}. The integral in the transfer operator, Eq. \ref{Seq:T}, is computed numerically by discretizing the length range of $l$ uniformly into $M_l$ intervals, and the angular range of $\omega$ uniformly into $M_\omega$ intervals.  Next, the integrals in Eqs \ref{Seq:Wi+1} and \ref{Seq:Pi} are computed numerically by discretizing the angular range of $\theta$ uniformly into $M_\theta$ intervals. In the general case of deformable bond angles, we found that taking $M_{\omega}=100$, $M_l = 50(l_{max}-l_{min})/l_e$ and $M_{\theta}=50$ is sufficient to achieve convergence. Here $l_{max}$ and $l_{min}$ represent the maximum and minimum bond length used in the numerical calculation, set as $2l_e$ and $0$, respectively. Note that increasing $l_{max}$ can cause numerical overflow for large applied force (e.g. $fl_e/k_BT>10^{2.5}$). Finally, the Gibbs free energy is calculated using Eq. \ref{Seq:G}, and the chain extension for a given force is calculated using Eq. \ref{Seq:rf}, where the derivative is approximated by finite differences. 

\begin{figure}
    \includegraphics{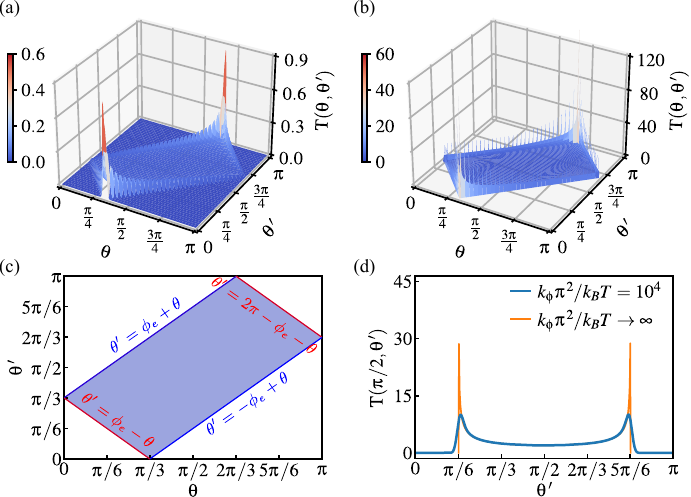}
    \caption{\label{sfig:TransferOperator}Transfer operator $\textrm T(\theta,\theta')$ for (a) a deformable bond angle ($\phi_e=60^\circ$, $k_\phi\pi^2/k_BT=10^4$) and (b) a fixed bond angle ($\phi_e=60^\circ$, $k_\phi\pi^2/k_BT \rightarrow \infty$). In both cases, the bonds are rigid, and no external force is applied. (c) shows the domain of the transfer operator in the case of fixed bond angle. (d) compares $\textrm{T}(\pi/2,\theta')$ for cases (a) and (b). In the deformable bond angle case, the Boltzmann factor $e^{-(k_\phi/2k_BT)(\phi-\phi_e)^2}$ equals $1$ at $\phi=\phi_e$, while in the fixed bond angle case, the delta function $\delta(\phi-\phi_e)$ is infinite at $\phi=\phi_e$. To enable a direct comparison, the transfer operators are normalized by their integrals over $\phi$, respectively given by $\int{e^{-(k_\phi/2k_BT)(\phi-\phi_e)^2}}\,d\phi=\sqrt{2\pi k_BT/k_\phi}$ and $\int{\delta(\phi-\phi_e)}\,d\phi=1$.}
\end{figure}

Fig. \ref{sfig:TransferOperator}a shows the transfer operator $\textrm{T}(\theta,\theta')$ for rigid bonds ($k_l l^2_e/k_B T \rightarrow\infty$) and deformable bond angles ($\phi_e=60^\circ$, $k_\phi\pi^2/k_BT=10^4$). Because the Boltzmann factor $e^{-k_BT(\phi-\phi_e)^2}$ strongly favors $\phi\approx\phi_e$, $\textrm{T}(\theta,\theta')$ shows sharp peaks in regions where this condition is satisfied. For example, when $\theta=\phi_e$, $\textrm{T}(\theta,\theta')$ is sharply peaked near $\theta'=0$, while at $\theta=\pi/2$, the peaks appear around $\theta'=\pi/2\pm\phi_e$ (Fig. \ref{sfig:TransferOperator}d). As $k_\phi$ increases, these peaks grow in magnitude. In addition, $\textrm{T}(\theta,\theta')$ is modulated by the geometric weighting $\sin{\theta}$, which reflects how orientations are distributed on the unit sphere. Near $\theta=0$ or $\theta=\pi$, $\sin{\theta}$ is small, so even if the Boltzmann factor is large, its product with $\sin{\theta}$ remains small. Conversely, around $\theta=\pi/2$, $\sin{\theta}$ is the largest, favoring those configurations unless the bending energy strongly disfavors them. 

\subsection{S2.2 Limiting Case 1: Freely Jointed Bonds}
When $k_\phi=0$, the bonds are freely jointed, and the transfer operator (Eq. \ref{Seq:T}) simplifies to:
\begin{equation}
    \textrm{T}\left(\theta\right) = 2\pi\int e^{-\frac{1}{k_BT}\left[v_{str}{\left(l\right)} - fl\cos{\theta}\right]} l^2\sin{\theta}\,dl
    \label{seq:T_FJB}
\end{equation}
Then, it is readily seen from Eqs \ref{Seq:B1} and \ref{Seq:Wi+1} that $W_{i+1}(\theta) = W_1(\theta)= \textrm{T}\left(\theta\right)$. Consequently, the partition function becomes:
\begin{equation}
     Z\left(f,T\right) \propto \left[2\pi\iint{e^{-\frac{1}{k_BT}\left[v_{str}{\left(l\right)} - fl\cos{\theta}\right]} l^2\sin{\theta}\,dl\,d\theta}\right]^N 
\end{equation}
Integration over $\theta$ can be carried out analytically, giving:
\begin{equation}
     Z\left(f,T\right) \propto \left[4\pi\int{e^{-\frac{1}{k_BT}\left[v_{str}{\left(l\right)}\right]}\frac{\sinh{(fl/k_BT)}}{fl/k_BT} l^2\,dl}\right]^N
     \label{seq:Z_FJB}
\end{equation}
The chain end-to-end distance $r(f,T)$ is then expressed as:
\begin{equation}
     r\left(f,T\right) = Nk_BT\frac{\partial}{\partial f}\ln{\left[\int{e^{-\frac{1}{k_BT}\left[v_{str}{\left(l\right)}\right]}\frac{\sinh{(fl/k_BT)}}{fl/k_BT} l^2\,dl}\right]}
     \label{seq:rf_FJB}
\end{equation}
These results agree with the findings in Ref. \cite{manca2012elasticity}.

When $k_l l^2_e/k_B T \rightarrow\infty$, bonds are rigid with $l=l_e$. The classical Freely Jointed Chain (FJC) model is then recovered \cite{rubinstein2003polymer}:
\begin{equation}
     r\left(f,T\right) = Nk_BT\frac{\partial}{\partial f}\ln{\left[\frac{\sinh{(fl_e/k_BT)}}{fl_e/k_BT}\right]} = Nl_e\mathcal{L}\left(\frac{fl_e}{k_BT}\right)
\end{equation}
where $\mathcal{L}(x)=\coth{(x)}-1/x$ is the Langevin function.

\subsection{S2.3 Limiting Case 2: Fixed Bond Angles}
When $k_\phi \pi^2/k_B T \rightarrow\infty$, the angle between two consecutive bonds is constant with $\phi=\phi_e$. The transfer operator (Eq. \ref{Seq:T}) becomes:
\begin{equation}
    \textrm{T}\left(\theta,\theta'\right) = \iint e^{-\frac{1}{k_BT}\left[v_{str}{\left(l\right)} - fl\cos{\theta}\right]} \delta(\phi-\phi_e)l^2\sin{\theta}\,d\omega\,dl
\end{equation}
where $\phi=\arccos[\sin \theta \sin \theta' \cos \omega + \cos \theta \cos \theta']$. Using the Dirac delta function composition rule, for fixed $\theta$ and $\theta'$, the delta function $\delta(\phi-\phi_e)$ can be expressed in terms of $\omega$ as:
\begin{equation}
    \delta(\phi-\phi_e) = \sum_i\frac{\delta(\omega-\omega_i)}{\abs{({d\phi}/{d\omega})_{\omega=\omega_i}}}
\end{equation}
where $\omega_i$ are the roots of $\phi(\omega)-\phi_e=0$. They satisfy
\begin{equation}\label{Seq:root}
    {\sin{\theta}\sin{\theta'}}\cos{\omega} + \cos{\theta}\cos{\theta'} - \cos{\phi_e} = 0 
\end{equation}
Several cases should be discussed:
\begin{enumerate}[(i)]
    \item If $\sin{\theta'}=0$ ($\theta'=0$ or $\theta'=\pi$), $\omega$ can be any value in the range of $[0,2\pi)$. According to Eq. \ref{Seq:root}, we have $\theta=\phi_e$ or $\theta=\pi-\phi_e$ depending on the value of $\theta'$ (Fig. \ref{sfig:Schematic_of_bond_angle}b). The transfer operators corresponding to these two special cases are given by:
    \begin{equation}
        \textrm{T}\left(\phi_e,0\right) = 2\pi\int{e^{-\frac{1}{k_BT}\left[v_{str}{\left(l\right)} - fl\cos{\phi_e}\right]}l^2\sin{\phi_e}\,dl}
    \end{equation}
    and: 
    \begin{equation}
    \textrm{T}\left(\pi-\phi_e,\pi\right) = 2\pi\int{e^{-\frac{1}{k_BT}\left[v_{str}{\left(l\right)} + fl\cos{\phi_e}\right]}l^2\sin{\phi_e}\,dl}
    \end{equation}
    For other values of $\theta$, we set $\textrm{T}\left(\theta,0\right)=0$ and $\textrm{T}\left(\theta,\pi\right)=0$. 
    
    \item If $\sin{\theta}=0$ ($\theta=0$ or $\theta=\pi$), $\omega$ can be any value in the range of $[0,2\pi)$ and $\theta'=\phi_e$ or $\theta'=\pi-\phi_e$ (Fig. \ref{sfig:Schematic_of_bond_angle}c). Then:
    \begin{equation}
        \textrm{T}\left(0,\phi_e\right) = 0
    \end{equation}
    and: 
    \begin{equation}
        \textrm{T}\left(\pi,\pi-\phi_e\right) = 0
    \end{equation}

    \item If $\sin{\theta}\sin{\theta'}\neq0$, then $\omega$ should satisfy
    \begin{equation}\label{Seq_g}
        \cos{\omega} = \frac{\cos{\phi_e}-\cos{\theta}\cos{\theta'}}{\sin{\theta}\sin{\theta'}} = g(\theta,\theta')
    \end{equation}

    \begin{enumerate}[(a)]
    \item If $\abs{g(\theta,\theta')}<1$, there are two possible values of $\omega$ in the range of $[0,2\pi)$: $\omega_1=\arccos{[g(\theta,\theta')]}$ and $\omega_2=2\pi-\arccos{[g(\theta,\theta')]}$ (Fig. \ref{sfig:Schematic_of_bond_angle}d). We also have:
    \begin{equation}
        \frac{d\phi}{d\omega} = \frac{\sin\theta\sin\theta'\sin{\omega}}{\sin{\phi}}
    \end{equation}
    Then:
    \begin{equation}
        \abs{\left(\frac{d\phi}{d\omega}\right)_{\omega=\omega_1}} = \abs{\left(\frac{d\phi}{d\omega}\right)_{\omega=\omega_2}} = \frac{\sin\theta\sin\theta'}{\sin{\phi_e}}\sqrt{1-\left(\frac{\cos{\phi_e}-\cos{\theta}\cos{\theta'}}{\sin{\theta}\sin{\theta'}}\right)^2}
    \end{equation}
    The delta function $\delta(\phi-\phi_e)$ becomes
    \begin{equation}
        \delta(\phi-\phi_e) = \sum_{i=1}^2\frac{\sin{\phi_e}}{\sqrt{\left(\sin{\theta}\sin{\theta'}\right)^2-\left(\cos{\phi_e}-\cos{\theta}\cos{\theta'}\right)^2}}\delta(\omega-\omega_i)
    \end{equation}
    Substituting it back to the transfer operator, we obtain:
    \begin{equation}
        \textrm{T}\left(\theta,\theta'\right) = \int e^{-\frac{1}{k_BT}\left[v_{str}{\left(l\right)} - fl\cos{\theta}\right]} \frac{2l^2\sin{\phi_e}\sin{\theta}}{\sqrt{\left(\sin{\theta}\sin{\theta'}\right)^2-\left(\cos{\phi_e}-\cos{\theta}\cos{\theta'}\right)^2}}\,dl
        \label{seq:T_FBA}
    \end{equation}
    
    \item If $\abs{g(\theta,\theta')}=1$, $\omega$ also has two possible values: $\omega_1=0$ and $\omega_2=\pi$ (Fig. \ref{sfig:Schematic_of_bond_angle}e). At these points, $({d\phi}/{d\omega})_{\omega=\omega_1}=({d\phi}/{d\omega})_{\omega=\omega_2}=0$, indicating a second-order zero of $\phi(\omega)-\phi_e$. For a second-order zero at $\omega=\omega_i$, the delta function generalization is:
    \begin{equation}
        \delta(\phi-\phi_e) = \sum_{i=1}^2\frac{\delta'(\omega-\omega_i)}{\abs{({d^2\phi}/{d\omega^2})_{\omega=\omega_i}}/2} = \sum_{i=1}^2\frac{2\sin{\phi_e}\delta'(\omega-\omega_i)}{\sin{\theta}\sin{\theta'}}
    \end{equation}
    and it immediately follows from the properties of the delta function derivative that $\textrm{T}\left(\theta,\theta'\right) = 0$.
    
    \item Finally, if $\abs{g(\theta,\theta')}>1$, there is no valid value of $\omega$ in the range of $[0,2\pi)$, so $\textrm{T}\left(\theta,\theta'\right) = 0$.
    \end{enumerate}
\end{enumerate}

If in addition $k_l l^2_e/k_B T \rightarrow\infty$ (rigid bonds), the Freely Rotating Chain (FRC) model is recovered. A statistical model for the FRC model was previously developed by Livadaru et al. \cite{livadaru2003stretching} using the TM technique. In their approach, the constraint of fixed bond angles was imposed by expressing the delta function $\delta(\phi-\phi_e)$ in terms of $\theta'$ and constructing the corresponding transfer operator to solve the probability function $B_i(\theta)$ for each bond.

The transfer operator $\textrm{T}(\theta,\theta')$ is non-zero only at specific points: $(\phi_e,0)$ and $(\pi-\phi_e,\pi)$, or when the condition $\abs{g(\theta,\theta')}<1$ is satisfied. As a result, $\textrm{T}(\theta,\theta')$ exhibits discontinuities, as illustrated in Fig. \ref{sfig:TransferOperator}b. Fig. \ref{sfig:TransferOperator}c shows the domain of $\textrm{T}(\theta,\theta')$, highlighting discontinuities at $\theta'=\phi_e\pm\theta$, $\theta'=-\phi_e+\theta$, and $\theta'=2\pi-\phi_e-\theta$. Fig. \ref{sfig:TransferOperator}d compares the transfer operators in the cases of fixed bond angle ($\phi_e=60^\circ, k_\phi \pi^2/k_B T \rightarrow\infty$) and deformable bond angle ($\phi_e=60^\circ$, $k_\phi\pi^2/k_BT=10^4$) when $\theta=\pi/2$. It is obvious that in the fixed bond angle case, discontinuities appear at $\pi/2\pm\phi_e$. Because of these discontinuities, the integral $\int {P_i(\theta')\textrm{T}(\theta,\theta')\,d\theta'}$ used in the computation of $W_{i+1}(\theta)$ must be handled with care. First, we identify the angles $\theta'_{r1}$ and $\theta'_{r2}$ where $\abs{g(\theta,\theta')}=1$ for each $\theta$. We then discretize the angular range $(\theta'_{r1},\theta'_{r2})$ uniformly into $M_\theta$ intervals, producing a set of points $\theta'_k$. At each point $\theta'_k$, $\textrm{T}(\theta,\theta'_k)$ is evaluated using Eq. \ref{seq:T_FBA}, and $P_i(\theta'_k)$ is interpolated from $P_i(\theta)$. In this case, we find that taking $M_{\theta}=250$ is sufficient to achieve convergence.

\section{S3 Bond Orientation Probabilities}
We examine the bond orientation probabilities $P_i(\theta)$ (probability density of finding the $i$th bond at an angle $\theta$) in several cases, highlighting the roles of the applied force $f$, bond stretching stiffness $k_l$, and bond bending stiffness $k_\phi$.

\subsection{S3.1 Freely jointed bonds}
\begin{figure}
    \includegraphics{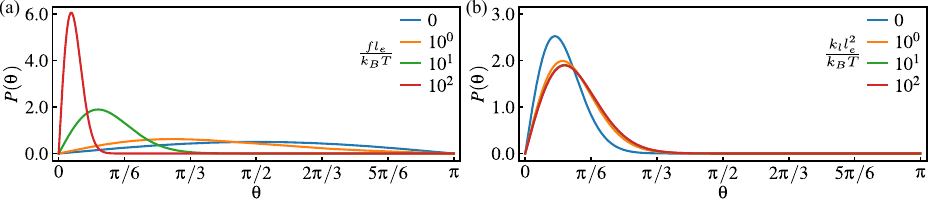}
    \caption{\label{sfig:Pdistribution_FJB}Bond orientation probability $P(\theta)$ for chains with freely jointed bonds. (a) Chains with rigid bonds ($k_l l^2_e/k_B T \rightarrow\infty$) under applied forces $fl_e/k_BT=0$, $10^0$, $10^1$, and $10^2$. (b) Chains with extensible bonds ($k_l l_e^2/k_BT=10^1$, $10^2$, $10^3$, and $10^4$) under an applied force of $fl_e/k_BT=10$.}
\end{figure}

In the freely jointed bond limit ($k_{\phi}=0$, Sec. S2.2), all bonds have the same orientation probability, $P_i(\theta)=P(\theta)$. The applied force $f$ biases $P(\theta)$ towards alignment with the force direction ($\theta=0^{\circ}$). Under no force, the bonds adopt random orientations, with $P(\theta)=\frac{1}{2}\sin{\theta}$ reflecting intrinsic geometric weighting (Fig. \ref{sfig:Pdistribution_FJB}a). As $f$ increases, $P(\theta)$ sharpens towards $\theta=0^{\circ}$, indicating bond alignment in the force direction. Bond stretching, governed by $k_l$, further modulates this behavior. When $k_l$ is small, the bond length $l$ fluctuates significantly from $l_e$, broadening the integral over $l$ in $W(\theta)=\textrm{T}(\theta)$, given by Eq. \ref{seq:T_FJB}. The force term $-fl\cos{\theta}$ in the exponent favors longer bonds ($l>l_e$) when $\cos{\theta}>0$ (i.e., for bonds aligned with the force). As a result, small $k_l$ amplifies alignment, sharpening $P(\theta)$. In contrast, when $k_l$ is large, the bond length remains nearly fixed at $l_e$, and only angular fluctuations contribute to the configurations, and $P(\theta)$ approaches the rigid bond limit (Fig. \ref{sfig:Pdistribution_FJB}b).

\subsection{S3.2 Fixed bond angles}
\begin{figure}
    \includegraphics{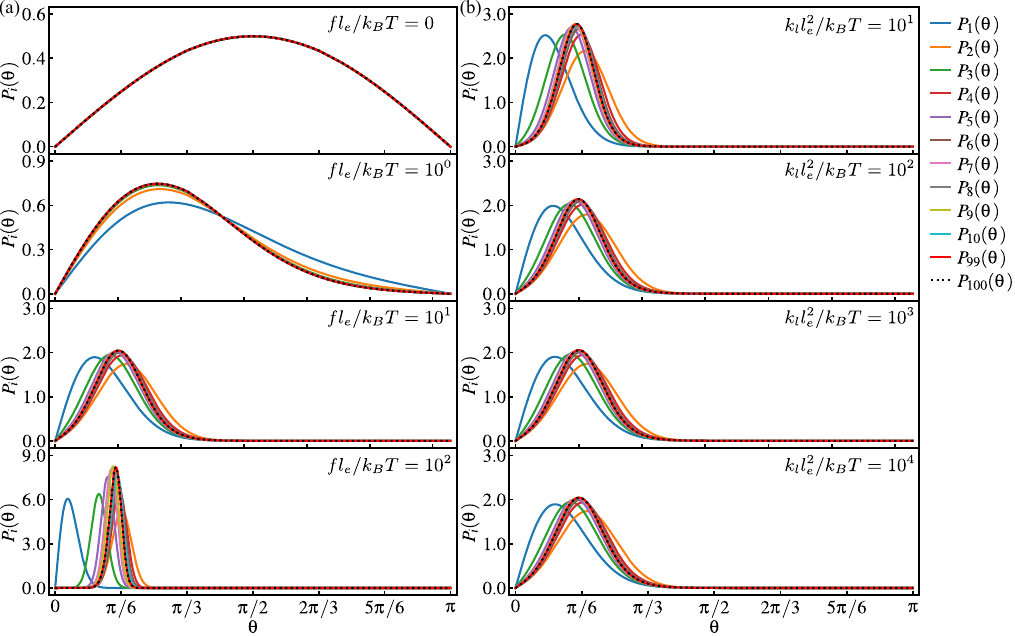}
    \caption{\label{sfig:Pdistribution_FBA_phi60}Bond orientation probability $P_i(\theta)$ for chains with a fixed bond angle $\phi_e=60^\circ$. (a) Chains with rigid bonds ($k_l l_e^2/ k_B T \rightarrow\infty$) under applied forces $fl_e/k_BT=0$, $10^0$, $10^1$, and $10^2$. (b) Chains with extensible bonds ($k_l l_e^2/k_BT=10^1$, $10^2$, $10^3$, and $10^4$) under an applied force of $fl_e/k_BT=10$.}
\end{figure}

In the fixed bond angle limit ($k_{\phi} \pi^2/k_B T \rightarrow \infty$, Sec. S2.3), the coupling between the orientations of two successive bonds via the bending energy leads to distinct orientation probability $P_i(\theta)$ for each bond. The first bond, $P_1(\theta)$, behaves identically to the freely jointed bonds case. As the bond index $i$ increases, $P_i(\theta)$ converges to a stable distribution, $P_\infty(\theta)$ (Fig. \ref{sfig:Pdistribution_FBA_phi60}). The applied force $f$ influences both the convergence rate and the shape of $P_\infty(\theta)$. At small forces, $P_\infty(\theta)$ is similar to $P_1(\theta)$. As the applied force increases, $P_\infty(\theta)$ sharpens around $\phi_e/2$ and converges at higher index number (Fig. \ref{sfig:Pdistribution_FBA_phi60}a). In the limit $f l_e/k_B T \rightarrow \infty$, the chain is fully straightened, and $P_\infty(\theta)$ collapses to a delta function, $\delta(\theta-\phi_e/2)$, corresponding to the all-\textit{trans} (zig-zag) configuration. Bond stretching, controlled by $k_l$, affects $P_\infty(\theta)$ similar to the freely jointed bonds case: a small $k_l$ favors alignment, while a large $k_l$ hinders bond stretching, mimicking rigid bonds (Fig. \ref{sfig:Pdistribution_FBA_phi60}b).

\subsection{S3.3 Deformable bond angles}
\begin{figure}
    \includegraphics{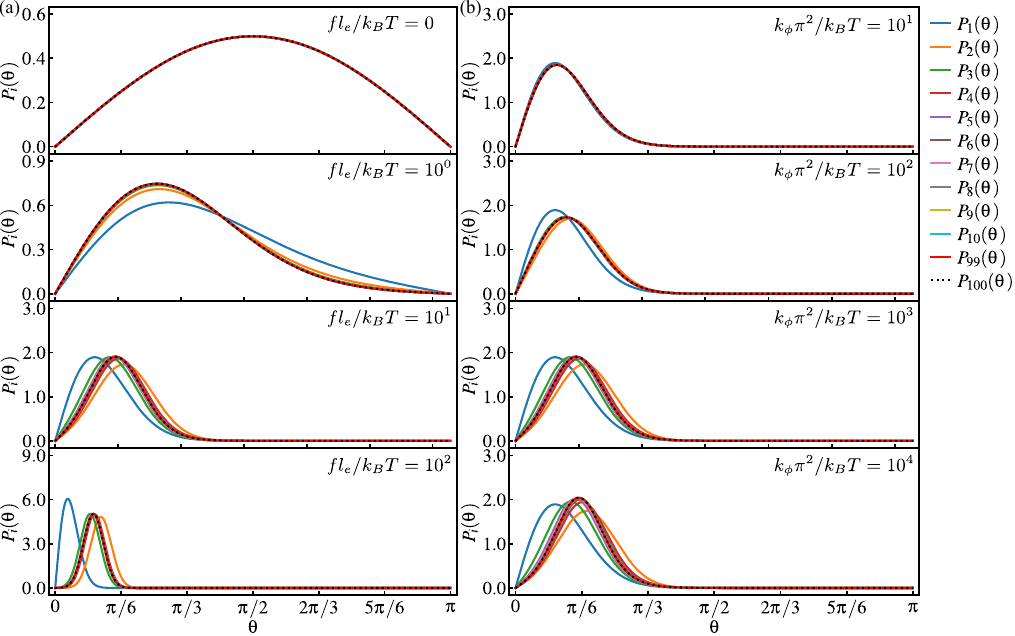}
    \caption{\label{sfig:Pdistribution_FRB_phi60}Bond orientation probability $P_i(\theta)$ for chains with rigid bonds ($k_l l^2_e/k_B T\rightarrow\infty$) and deformable bond angles ($\phi_e=60^\circ$). (a) Chains with deformable bond angles ($k_\phi\pi^2/k_BT=10^3$) under applied forces $fl_e/k_BT=0$, $10^0$, $10^1$, and $10^2$. (b) Chain with deformable bond angles ($k_\phi\pi^2/k_BT=10^1$, $10^2$, $10^3$, and $10^4$) under an applied force of $fl_e/k_BT=10$.}
\end{figure}

In the general case of deformable bond angles (Sec. S2.1), each bond exhibits a distinct orientation probability $P_i(\theta)$. The first bond, $P_1(\theta)$, behaves identically to the freely jointed bonds case. For increasing bond index number $i$, $P_i(\theta)$ converges to a stable distribution, $P_\infty(\theta)$ (Fig. \ref{sfig:Pdistribution_FRB_phi60}). Both the applied force $f$ and the bond bending stiffness $k_\phi$ affect the convergence rate and the shape of $P_\infty(\theta)$. At small forces, $P_\infty(\theta)$ remains similar to $P_1(\theta)$; at large forces, it sharpens around $\phi_e/2$, requiring more iterations to converge (Fig. \ref{sfig:Pdistribution_FRB_phi60}a).
The stiffness $k_\phi$ also plays a critical role: small $k_\phi$ allows easy bond angle deformation, making $P_\infty(\theta)$ similar to the freely jointed bonds case, while large $k_\phi$ causes $P_i(\theta)$ to approach the fixed bond angle limit (Fig. \ref{sfig:Pdistribution_FRB_phi60}b).

\section{S4. Force-Extension Relationships}
\subsection{S4.1 Effect of chain length}
For freely jointed bonds ($k_{\phi}=0$), the Gibbs partition function of a chain with $N$ bonds can be expressed as the partition function of a single bond raised to the power $N$ (Eq. \ref{seq:Z_FJB}). Consequently, the force-extension curve under prescribed force is independent of the chain length $N$, consistent with the findings in Ref. \cite{manca2012elasticity}.

When bond angles are constrained, the Gibbs partition function cannot be factorized, and the force-extension curve depends on $N$. However, for sufficiently large $N$, the Gibbs partition function can be approximated as $Z \propto \left[\int W_{\infty}(\theta)\,d\theta\right]^N$, where $W_{\infty}(\theta) = \int {P_{\infty}\left(\theta'\right)\textrm{T}\left(\theta,\theta'\right)\,d\theta'}$ and $P_\infty(\theta)$ is the converged bond orientation probability. In this regime, the force-extension curve becomes independent on $N$, corresponding to the thermodynamic limit. 

\begin{figure}
    \includegraphics{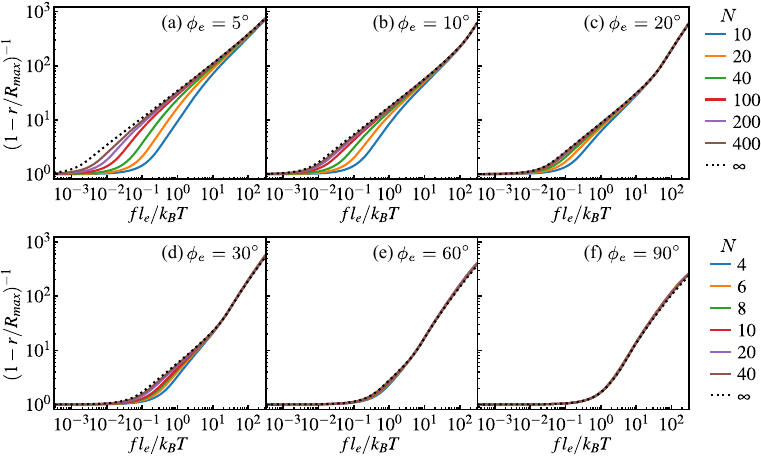}
    \caption{\label{sfig:FBA_Dependence_on_N}Force-extension curves for chains with rigid bonds and fixed bond angles (FRC), showing dependence on chain length $N$ for varying bond angles $\phi_e$. (a) $\phi_e=5^\circ$. (b) $\phi_e=10^\circ$. (c) $\phi_e=20^\circ$. (d) $\phi_e=30^\circ$. (e) $\phi_e=60^\circ$. (f) $\phi_e=90^\circ$. Here $R_{max}$ is the contour length of the chain, $R_{max}=Nl_e\cos{(\phi_e/2)}$.} 
\end{figure}

For chains with fixed bond angles at $\phi_e$ ($k_{\phi} \pi^2/k_B T \rightarrow\infty$), the chain length $N$ required to achieve this thermodynamic limit depends on $\phi_e$, as shown in Fig. \ref{sfig:FBA_Dependence_on_N}. For large bond angles ($\phi_e\geq30^\circ$), a small $N$ ($N=40$) is sufficient (Fig. \ref{sfig:FBA_Dependence_on_N}d-f). However, for smaller bond angles ($\phi_e\rightarrow0^{\circ}$), a larger $N$ is needed to reach size independence (Fig. \ref{sfig:FBA_Dependence_on_N}a-c). These results are consistent with the findings in Ref. \cite{livadaru2003stretching}.

\begin{figure}
    \includegraphics{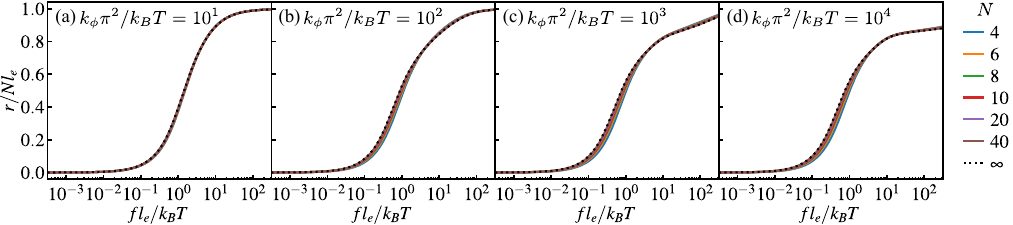}
    \caption{\label{sfig:FRB_Dependence_on_N}Force-extension curves for chains with rigid bonds and deformable bond angles ($\phi_e=60^\circ$), showing dependence on chain length $N$ for varying bond bending stiffness $k_\phi$. (a) $k_\phi\pi^2/k_BT=10^1$. (b) $k_\phi\pi^2/k_BT=10^2$. (c) $k_\phi\pi^2/k_BT=10^3$. (d) $k_\phi\pi^2/k_BT=10^4$.}
\end{figure}

For chains with deformable bond angles, the effect of $N$ on the force-extension curve is modulated by the bond bending stiffness $k_\phi$, as shown in Fig. \ref{sfig:FRB_Dependence_on_N}. When $k_\phi$ is small, the chain behaves similarly to the freely jointed bonds case, with the force-extension curve nearly independent of $N$ (Fig. \ref{sfig:FRB_Dependence_on_N}a). As $k_\phi$ increases, bond angles stiffen, and geometrical constraints between adjacent bonds become more pronounced. This increases the dependence on $N$, requiring larger chain lengths to reach the thermodynamic limit (Fig. \ref{sfig:FRB_Dependence_on_N}b-d).

\subsection{S4.2 Effect of bond angle deformation}
\begin{figure}
    \includegraphics{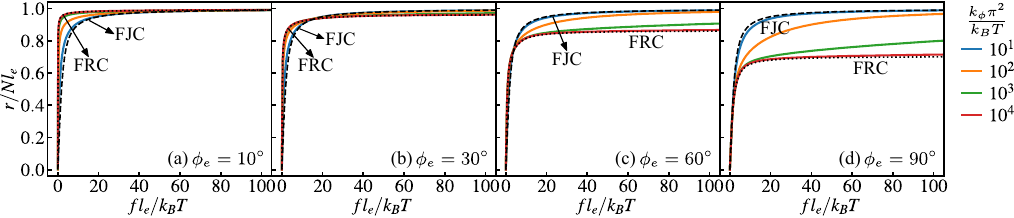}
    \caption{\label{sfig:Dependence_on_kphi}Dependence of the force-extension curve on the bond bending stiffness $k_\phi$. The bonds are rigid ($k_l l^2_e/k_B T \rightarrow \infty$), with equilibrium bond angles of (a) $\phi_e=10^\circ$, (b) $\phi_e=30^\circ$, (c) $\phi_e=60^\circ$, and (d) $\phi_e=90^\circ$.}
\end{figure}
We examine the effect of bond angle deformation on the stretching response of chains with rigid bonds ($k_l l^2_e/k_B T\rightarrow \infty$). The bond bending stiffness $k_\phi$ governs the transition between the FJC and FRC models. When $k_\phi=0$, the FJC model is recovered. As $k_\phi \pi^2/k_B T \rightarrow\infty$, the FRC model is recovered, with bond angles fixed at $\phi_e$. Fig.~\ref{sfig:Dependence_on_kphi} shows the force-extension behavior for varying stiffness $k_\phi$ and equilibrium angle $\phi_e$. At low forces, the FRC model predicts a softer response compared to the FJC model. This occurs because the fixed bond angle in the FRC reduces the number of possible available conformations, favoring larger extensions under small force. At higher forces, however, this behavior reverses: the FRC model becomes stiffer because the fixed bond angles restrict the chain ability to extend. As the bonds align with the direction of the applied force, the FRC approaches its contour length, $R_{max}=Nl_e\cos(\phi_e/2)$. In contrast, the FJC, free from these angular constraints, can reach a larger contour length $R_{max}=Nl_e$. The equilibrium bond angle $\phi_e$ at a given $k_{\phi}$ also impacts the chain behavior, primarily by setting the contour length of the chain. In particular, for a FRC, the contour length ranges from $R_{max}=N l_e$ (the FJC contour length) to $R_{max}=Nl_e/\sqrt{2}$ when $\phi_e$ goes from $0^{\circ}$ to $90^{\circ}$. 

\section{S5 A new explicit formula for the FRC model}
To evaluate the entropic energy $\Psi_{ent}$ in the proposed dFRC model, a closed-form expression for the force-extension response of the FRC is required. Livadaru et al. \cite{livadaru2003stretching} identified three distinct force-extension regimes for the FRC using scaling analysis. These authors further proposed a global formula for the FRC, based on the the interpolation formula for the Worm-Like Chain (WLC) model by Marko and Siggia \cite{marko1995stretching}. However, their formulation requires the numerical inversion of the WLC interpolation formula and involves a fitting parameter, making it unsuitable for our purpose. In the following, we propose a new formula for the FRC, which is explicit, valid across the whole force range, and does not require any fitting parameter.  

\begin{figure}
    \includegraphics{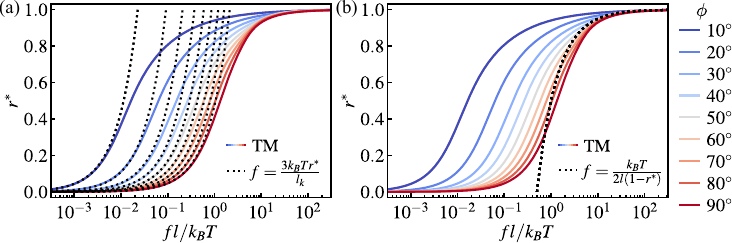}
    \caption{\label{sfig:TestSimplifiedRelation}
    Comparison between the force-extension predictions obtained from the transfer matrix (TM) method and the expressions governing the limiting behaviors for a freely rotating chain (FRC) with fixed bond length $l$ and bond angle $\phi$. 
    (a) In the small deformation regime ($r^* \rightarrow 0$), the force approximates that of a Gaussian chain: $f = \frac{3k_BT r^*}{l_k}$. 
    (b) In the large deformation regime ($r^* \rightarrow 1$), the force follows the divergence form: $f = \frac{k_BT}{2l(1 - r^*)}$.}
\end{figure}

To this end, we first compute the reference force-extension curves using the TM method for a range of fixed bond angles $\phi$ (Fig. \ref{sfig:TestSimplifiedRelation}). From these results, we confirm two key limiting behaviors previously identified by Livadaru et al. \cite{livadaru2003stretching}:
\begin{enumerate}[(i)]
\item In the small deformation regime ($r^*\rightarrow0$), the FRC behaves like a Gaussian chain, with the force given by
\begin{equation}
    f = \frac{3k_BTr^*}{l_k}
\end{equation}
as shown in Fig. \ref{sfig:TestSimplifiedRelation}a. Here $r^*=r/R_{max}$ represents the relative end-to-end distance, with the contour length given by $R_{max} = Nl\cos{\left(\phi/2\right)}$. The equivalent Kuhn length is $l_k =\left<r^2\right>/R_{max}= {2l\cos{\left(\phi/2\right)}}/(1-\cos{\phi})$, where the mean-square end-to-end distance is $\left<r^2\right>=Nl^2(1+\cos{\phi})/(1-\cos{\phi})$.
\item In the large deformation regime ($r^*\rightarrow1$), the force approximately diverges as
\begin{equation}\label{seq:f_large}
    f=\frac{k_BT}{2l(1-r^*)}
\end{equation}
as shown in Fig. \ref{sfig:TestSimplifiedRelation}b.
\end{enumerate}

To recover these two limiting behaviors, we start with the inverse Langevin function ($\beta=\mathcal{L}^{-1}(r^*)$ and $\mathcal{L}(x)=\coth{(x)-1/x}$) because it accurately reproduces the small-deformation limit ($\beta \approx 3r^*$ as $r^*\rightarrow 0$). However, this function does not correctly capture the large deformation behavior ($\beta \approx \frac{1}{1-r^*}$ as $r^*\rightarrow 1$). To correct this, we introduce an additional term that vanishes at small deformations, preserving the low-force limit while capturing the high-force behavior described by Eq. \ref{seq:f_large}. The specific form of Eq. 8 is obtained by informed trial-and-error, to simultaneously recover the correct limiting behaviors while achieving good agreement with the reference TM curves for different bond angles. 

\begin{figure}
   \includegraphics{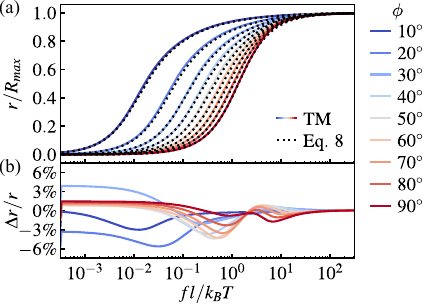}
   \caption{Comparison of the TM calculations and Eq. 8 for a FRC with fixed bond length $l$ and bond angle $\phi$. (a) Force-extension curves. (b) Relative error of Eq. 8 compared to TM calculations.} 
   \label{fig:SimplifiedRelation}
\end{figure}

We verify the accuracy of Eq. 8 by comparing its predictions to the TM calculations across a wide range of bond angles ($\phi=10^\circ-90^\circ$). As shown in Fig. \ref{fig:SimplifiedRelation}, Eq. 8 shows a remarkable agreement with TM calculations, maintaining a relative error below $6\%$ over the entire force range ($fl_e/k_BT=10^{-3.5}-10^{2.5}$). To the best of our knowledge, this is the first global explicit formula for the FRC model that is entirely free of fitting parameters. Additionally, in the special case of a freely jointed chain ($l_k=l$ and $R_{max}=Nl$), the correction term becomes zero, and Eq. 8 correctly reduces to the standard FJC model, $f=\frac{k_BT}{l_k}\beta$.

\section{S6 Validation of the semi-analytical model}
The proposed semi-analytical dFRC model assumes that the behavior of a chain with fluctuating bond lengths and bond angles can be approximated by a FRC with uniform and non-fluctuating bond length and bond angle, determined such that they minimize the Helmholtz free energy of the chain at a prescribed chain stretch. In this section, we assess the validity of these assumptions by analyzing the probability distributions of bond length and bond angle from TM calculations and comparing their most probable values to those obtained using the semi-analytical model. Additionally, we validate the semi-analytical model by comparing predicted force-extension curves to reference curves obtained from TM claculations. Three representative cases are discussed.

\subsection{S6.1 Freely jointed extensible bonds}
We first consider the case of a chain with freely jointed bonds ($k_\phi=0$). At an applied force $f$, the transfer operator simplifies to $\textrm{T}\left(\theta\right)$ (Eq. \ref{seq:T_FJB}). In this scenario, all bonds have the same length probability distribution, $P^l(l)$, which is given by
\begin{equation}
    P^l(l) = \frac{\int e^{-\frac{1}{k_BT}\left[v_{str}(l) - fl\cos{\theta}\right]}\,d\theta}{\iint e^{-\frac{1}{k_BT}\left[v_{str}(l) - fl\cos{\theta}\right]}\,d\theta\,dl}
\end{equation}
The average bond length is
\begin{equation}
    \langle l \rangle = \int lP^l(l)\,dl
\end{equation}
and the corresponding variance is
\begin{equation}
    \sigma_l^2 = \int (l - \langle l \rangle)^2P^l(l)\,dl
\end{equation}

\begin{figure}
    \includegraphics{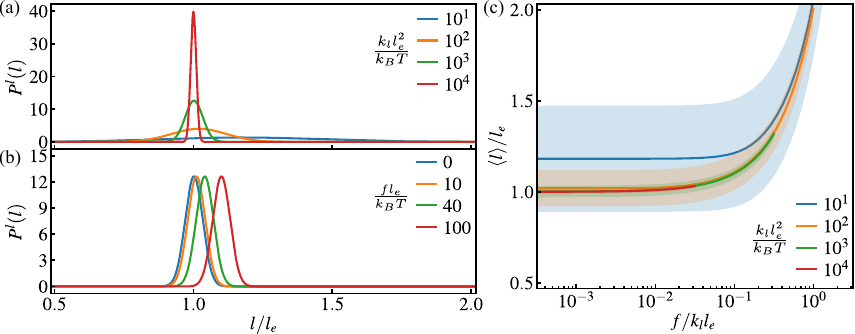}
    \caption{\label{fig:BondLengthDistribution}Bond length probability distribution for a chain with freely jointed extensible bonds. (a) Bond length probability at zero applied force ($f=0$) with varying bond stretching stiffness: $k_l l_e^2/k_BT=10^1$, $10^2$, $10^3$, and $10^4$. (b) Bond length probability for bond stretching stiffness $k_l l_e^2/k_BT=10^3$, under different applied forces: $fl_e/k_BT=0$, $10$, $40$, and $100$. (c) Average bond length and standard deviation of the bond length as a function of $f/k_l l_e$.}
\end{figure}

The bond length probability distribution $P^l(l)$ is affected by two key factors: the bond stretching stiffness $k_l$ and the applied force $f$. The stiffness $k_l$ controls the spread of the distribution. As $k_l$ increases, $P^l(l)$ becomes more narrowly concentrated around its mean value, indicating reduced bond length variability (Fig. \ref{fig:BondLengthDistribution}a). In the limit $k_l l^2_e /k_B T\rightarrow\infty$, the bond length is fixed at $l_e$, and the $P^l(l)$ converges to a delta function, $P^l(l)=\delta(l-l_e)$. On the other hand, the force $f$ primarily affects the average bond length. Increasing $f$ shifts the peak of $P^l(l)$ to the right, leading to a higher average bond length (Fig. \ref{fig:BondLengthDistribution}b). Fig. \ref{fig:BondLengthDistribution}c illustrates the dependence of the average bond length $\langle l\rangle$ and the standard deviation $\sigma_l$ on the dimensionless force $f/k_ll_e$. As $f/k_ll_e$ increases, $\langle l\rangle$ increases while $\sigma_l$ decreases.

In the semi-analytical model, the Helmholtz free energy for a FJC with uniform and non-fluctuating bond length $l$ is
\begin{equation}\label{seq:energy_simplified_FJB}
    \Psi = Nv_{str}\left(l\right) + Nk_BT\left[\frac{r\beta}{Nl} + \ln{\left(\frac{\beta}{\sinh{\beta}}\right)}\right]
\end{equation}
where $\beta=\mathcal{L}^{-1}(r/Nl)$ is the inverse Langevin function. The force-extension relationship is given by
\begin{equation}
     f = \frac{k_BT}{l}\beta
\end{equation}
where the bond length $l$ is determined by minimizing the Helmholtz free energy at a prescribed chain end-to-end distance $r$. This extensible FJC (eFJC) model was first proposed by Mao et al. \cite{mao2017rupture}, who used a logarithmic strain-based stretching energy $v_{str}(l)=\frac{1}{2}k_l[\ln{(l/l_e)}]^2$. The simpler quadratic form $v_{str}(l)=\frac{1}{2}k_l(l-l_e)^2$ was considered in their subsequent work \cite{talamini2018progressive}. To enhance computational efficiency, Li and Bouklas proposed an approximate solution for $l$ using the roots of cubic equations via trigonometric functions \cite{li2020variational}. Although this eFJC model has been widely used in the multiscale modeling of polymer fracture \cite{mulderrig2021affine,arunachala2023multiscale,arunachala2024multiscale}, its assumption of uniform and non-fluctuating bond length has not been validated, nor have its force-extension predictions been validated, leaving its accuracy uncertain. 
\begin{figure}
    \includegraphics{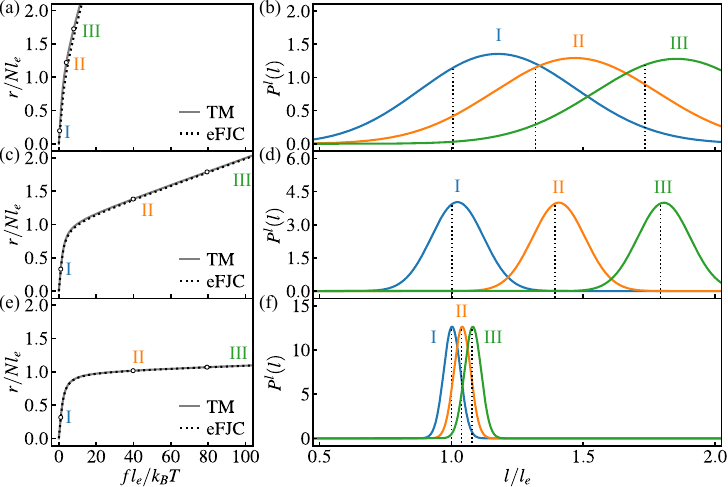}
    \caption{\label{fig:FJB_ComparisonStatisticalAndSimplified}Comparison of statistical (TM) and eFJC models for chains with freely jointed extensible bonds. (a), (c), and (e): Force-extension curves for bond stretching stiffness: $k_ll_e^2/k_BT=10^1$, $10^2$, and $10^3$, respectively. (b), (d), and (f): Corresponding bond length probability distributions in the statistical model at different applied forces, where the black dotted lines represent the corresponding optimized bond lengths in the eFJC model.}
\end{figure}

We evaluate the accuracy of the eFJC model by comparing its predictions to numerical results from the TM calculations. Notably, even at a relatively low bond stretching stiffness ($k_l l_e^2/k_BT=10^1$), the eFJC model still provides a good approximation of the chain force-extension response (Fig. \ref{fig:FJB_ComparisonStatisticalAndSimplified}a), despite the broad bond length distribution deviating significantly from the uniform bond length assumption (Fig. \ref{fig:FJB_ComparisonStatisticalAndSimplified}b). As $k_l l_e^2/k_BT$ increases, the accuracy of the eFJC model increases. For $k_ll_e^2/k_BT=10^3$, it accurately reproduces the chain stretching behavior (Fig. \ref{fig:FJB_ComparisonStatisticalAndSimplified}e), and the optimized bond length aligns with the most probable bond length in the bond length probability distribution obtained from TM calculations (Fig. \ref{fig:FJB_ComparisonStatisticalAndSimplified}f). 

\subsection{S6.2 Extensible bonds and fixed bond angles}
We next consider the case of a chain with extensible bonds and fixed bond angles ($k_\phi \pi^2/k_B T\rightarrow\infty$) (extensible FRC). The transfer operator becomes (Sec. S2.3)
\begin{equation}
    \textrm{T}\left(\theta,\theta'\right) = \begin{cases}
    2\pi\sin{\theta}\int{e^{-\frac{1}{k_BT}\left[v_{str}{\left(l\right)} - fl\cos{\theta}\right]}l^2\,dl} & (\theta,\theta')=(\phi_e,0) \textrm{ or } (\pi-\phi_e,\pi)\\
    0 & (\theta,\theta')=(0,\phi_e) \textrm{ or } (\pi,\pi-\phi_e)\\
    \frac{2\sin{\phi_e}\sin{\theta}\int e^{-\frac{1}{k_BT}\left[v_{str}{\left(l\right)} - fl\cos{\theta}\right]}l^2\,dl}{\sqrt{\left(\sin{\theta}\sin{\theta'}\right)^2-\left(\cos{\phi_e}-\cos{\theta}\cos{\theta'}\right)^2}} & \sin{\theta}\sin{\theta'}\neq0 \textrm{ and } \abs{g(\theta,\theta')}<1\\
    0 & \sin{\theta}\sin{\theta'}\neq0 \textrm{ and } \abs{g(\theta,\theta')}\geq1\\
    \end{cases}
\end{equation}
where $g(\theta,\theta')={(\cos{\phi_e}-\cos{\theta}\cos{\theta'})}/{\sin{\theta}\sin{\theta'}}$, Eq. \ref{Seq_g}. Now consider the transfer operator before the integration with respect to $l$:
\begin{equation}
    \textrm{T}^l\left(\theta,\theta'\right) = \begin{cases}
    2\pi\sin{\theta}{e^{-\frac{1}{k_BT}\left[v_{str}{\left(l\right)} - fl\cos{\theta}\right]}l^2} & (\theta,\theta')=(\phi_e,0) \textrm{ or } (\pi-\phi_e,\pi)\\
    0 & (\theta,\theta')=(0,\phi_e) \textrm{ or } (\pi,\pi-\phi_e)\\
    \frac{2\sin{\phi_e}\sin{\theta}e^{-\frac{1}{k_BT}\left[v_{str}{\left(l\right)} - fl\cos{\theta}\right]}l^2}{\sqrt{\left(\sin{\theta}\sin{\theta'}\right)^2-\left(\cos{\phi_e}-\cos{\theta}\cos{\theta'}\right)^2}} & \sin{\theta}\sin{\theta'}\neq0 \textrm{ and } \abs{g(\theta,\theta')}<1\\
    0 & \sin{\theta}\sin{\theta'}\neq0 \textrm{ and } \abs{g(\theta,\theta')}\geq1\\
    \end{cases}
\end{equation}
Under an applied force $f$, the bond length probability distribution of the $(i+1)$th bond, $P_{i+1}^l(l)$, is given by
\begin{equation} 
    P_{i+1}^l(l) = \frac{\iint{P_i(\theta')\textrm{T}^l\left(\theta,\theta'\right)\,d\theta'\,d\theta}}{\iiint{P_i(\theta')\textrm{T}^l\left(\theta,\theta'\right)\,d\theta'\,d\theta\,dl}}
\end{equation}
recalling that $P_i(\theta)$ represents the bond orientation probability of the $i$th bond. The bond length probability distribution of the $1$st bond is
\begin{equation} 
    P_1^l(l) = \frac{\int{e^{-\frac{1}{k_BT}\left[v_{str}{\left(l\right)} - fl\cos{\theta}\right]} l^2\sin{\theta}\,d\theta}}{\iint{e^{-\frac{1}{k_BT}\left[v_{str}{\left(l\right)} - fl\cos{\theta}\right]} l^2\sin{\theta}\,d\theta\,dl}}
\end{equation}

\begin{figure}
    \includegraphics{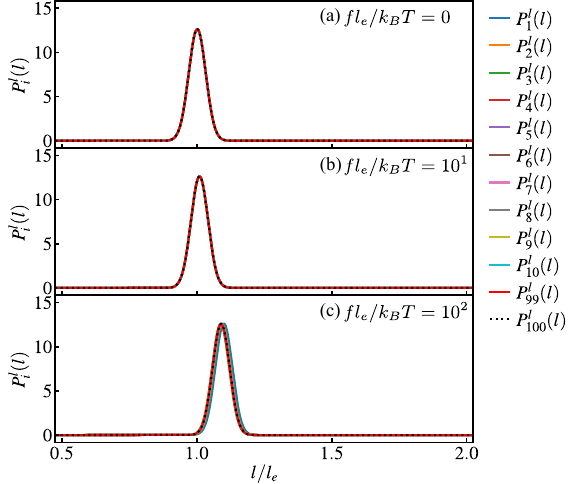}
    \caption{\label{sfig:FBA_Pli_distribution} Bond length probability density $P_{i}^l(l)$ for chains with extensible bonds ($k_ll_e^2/k_BT=10^3$) and a fixed bond angle of $\phi_e=60^\circ$, subjected to different applied forces: (a) $fl_e/k_BT=0$, (b) $fl_e/k_BT=10^1$, and (c) $fl_e/k_BT=10^2$.}
\end{figure}

As shown in Fig. \ref{sfig:FBA_Pli_distribution}a, at zero force, all bonds have the same bond length probability distribution $P_{i}^l(l)$. As the force increases, $P_{i}^l(l)$ varies from bond to bond but converges to a steady-state distribution $P_\infty^l(l)$ as the bond index $i$ increases. Although higher forces require a larger bond index to achieve convergence, the process remains relatively fast (Fig. \ref{sfig:FBA_Pli_distribution}c). Therefore, the converged distribution $P_\infty^l(l)$ is a suitable representation of the bond length probability distribution in the chain. In the following analysis, we focus on examining the mean and variance of $P_\infty^l(l)$ for different parameters. The average bond length is
\begin{equation}
    \langle l \rangle = \int lP_\infty^l(l)\,dl
\end{equation}
and the corresponding variance is
\begin{equation}
    \sigma_l^2 = \int (l - \langle l \rangle)^2P_\infty^l(l)\,dl
\end{equation}

\begin{figure}
    \includegraphics{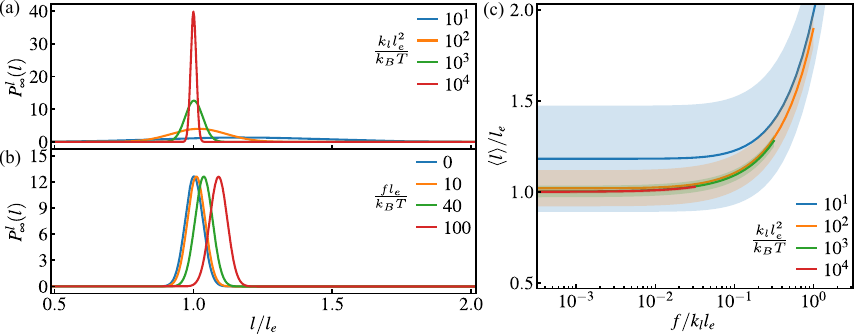}
    \caption{\label{sfig:BondLengthDistribution}Converged bond length probability distribution for a chain with extensible bonds and a fixed bond angle of $\phi_e=60^\circ$. (a) Probability distributions at zero applied force ($f=0$) for bond stretching stiffness: $k_ll_e^2/k_BT=10^1$, $10^2$, $10^3$, and $10^4$. (b) Probability distributions for a fixed bond stretching stiffness $k_l l_e^2/k_BT=10^3$ under different applied forces: $fl_e/k_BT=0$, $10$, $40$, and $100$. (c) Average bond length and standard deviation as a function of $f/k_ll_e$.}
\end{figure}

The probability distribution $P_\infty^l(l)$ for fixed bond angles exhibits similar trends to the freely jointed bonds case. The bond stretching stiffness $k_l$ controls the spread of $P_\infty^l(l)$, with higher $k_l$ values leading to a narrower distribution centered around the average bond length (Fig. \ref{sfig:BondLengthDistribution}a). In the limit $k_l l^2_e/k_B T \rightarrow\infty$, the distribution approaches $\delta(l-l_e)$. The applied force $f$ primarily affects the average bond length, with increasing $f$ shifting the peak of $P_\infty^l(l)$ to higher average bond length (Fig. \ref{sfig:BondLengthDistribution}b). Fig. \ref{sfig:BondLengthDistribution}c shows that $\langle l\rangle$ increases and $\sigma_l$ decreases as $f/k_ll_e$ increases, consistent with the freely jointed bonds case.

\begin{figure}
    \includegraphics{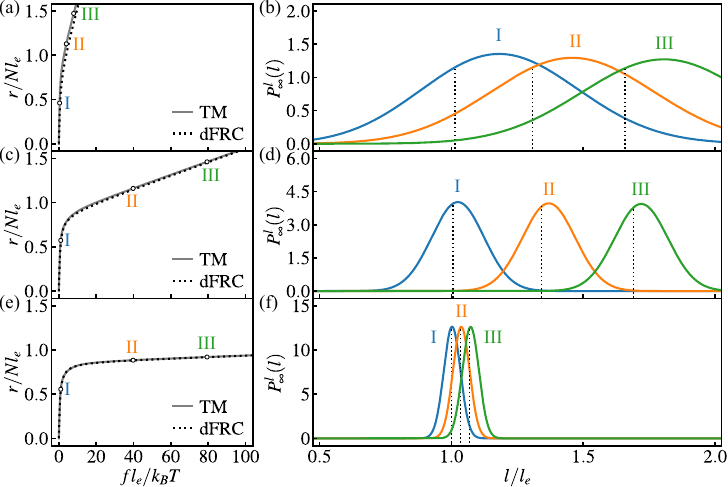}
    \caption{\label{sfig:FBA_ComparisonStatisticalAndSimplified}Comparison of statistical (TM) and dFRC models for chains with extensible bonds and a fixed bond angle of $\phi_e=60^\circ$. (a), (c), and (e): Force-extension curves for bond stretching stiffness $k_ll_e^2/k_BT=10^1$, $10^2$, and $10^3$, respectively. (b), (d), and (f): Corresponding converged bond length distributions in the statistical model at different applied forces, where the black dotted lines represent the corresponding optimized bond lengths in the dFRC model.}
\end{figure}

The comparison between the dFRC model and TM calculations for fixed bond angles also mirrors the freely jointed bonds case. At $k_ll_e^2/k_BT=10^1$, the dFRC model approximates the force-extension response obtained from the TM calculation well (Fig. \ref{sfig:FBA_ComparisonStatisticalAndSimplified}a), although the bond length distribution deviates from the uniform bond length assumption (Fig. \ref{sfig:FBA_ComparisonStatisticalAndSimplified}b). As $k_ll_e^2/k_BT$ increases, the dFRC model converges to the TM calculation, achieving excellent agreement at $k_ll_e^2/k_BT=10^3$ (Fig. \ref{sfig:FBA_ComparisonStatisticalAndSimplified}e). At this point, the optimized bond length in the dFRC model matches the most probable bond length in the distribution (Fig. \ref{sfig:FBA_ComparisonStatisticalAndSimplified}f).

\subsection{S6.3 Rigid bonds and deformable bond angles}
\begin{figure}
    \includegraphics{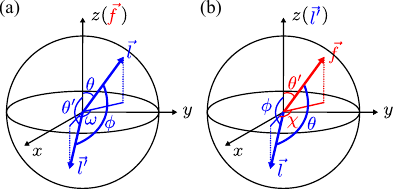}
    \caption{\label{sfig:Schematic_of_bond_angle_change}Schematic representation of coordinates used in the calculation of (a) $P_i(\theta)$ and (b) $P^\phi_i(\phi)$.}
\end{figure}

We next examine the bond angle probability distribution in a chain with rigid bonds ($k_l l^2_e/k_B T\rightarrow\infty$) and deformable bond angles. The transfer operator becomes
\begin{equation}
    \textrm{T}\left(\theta,\theta'\right) = \int e^{-\frac{1}{k_BT}\left[v_{ben}{\left(\phi\right)} - fl_e\cos{\theta}\right]}\sin{\theta}\,d\omega
\end{equation}
where $\theta$ and $\theta'$ are the polar angles of the $(i+1)$th and $i$th bond orientations, respectively, measured relative to the direction of the applied force $\vec{f}$ (Fig. \ref{sfig:Schematic_of_bond_angle_change}a). The bond angle between these two bonds is given by $\phi = \arccos{\left[\sin{\theta}\sin{\theta'}\cos{\omega} + \cos{\theta}\cos{\theta'}\right]}$, where $\omega$ is the difference in their azimuthal angle. The orientation probability of the $(i+1)$th bond is then given as $P_{i+1}(\theta)=\int{P_i(\theta')\textrm{T}\left(\theta,\theta'\right)\,d\theta'}/\iint{P_i(\theta')\textrm{T}\left(\theta,\theta'\right)\,d\theta'\,d\theta}$.

We now introduce the bond angle probability distribution $P^\phi_{i+1}(\phi)$, which is the probability of observing a bond angle $\phi$ between the $i$th and $(i+1)$th bonds. To express everything in terms of $\phi$ (instead of $\theta$), we rotate the coordinate system so that the new $z$-axis aligns with the $i$th bond $\vec{l}'$ (Fig. \ref{sfig:Schematic_of_bond_angle_change}b). In this new frame, $\theta'$ and $\phi$ become the polar angles of the applied force $\vec{f}$ and the $(i+1)$th bond $\vec{l}$ relative to $\vec{l}'$, respectively. Then the angle between $\vec{l}$ and $\vec{f}$, $\theta$, can be expressed as $\theta=\arccos{[\sin{\theta'}\sin{\phi}\cos{\chi} + \cos{\theta'}\cos{\phi}]}$, where $\chi$ is the difference in azimuthal angles between $\vec{f}$ and $\vec{l}$.

The bond angle probability distribution $P^\phi_{i+1}(\phi)$ is thus
\begin{equation}
    P_{i+1}^\phi(\phi) = \frac{\iint P_i(\theta')e^{-\frac{1}{k_BT}\left[v_{ben}{\left(\phi\right)} - fl_e\cos{\theta}\right]}\sin{\phi}\,d\chi\,d\theta'}{\iiint P_i(\theta')e^{-\frac{1}{k_BT}\left[v_{ben}{\left(\phi\right)} - fl_e\cos{\theta}\right]}\sin{\phi}\,d\chi\,d\theta'\,d\phi}
\end{equation}

\begin{figure}
    \includegraphics{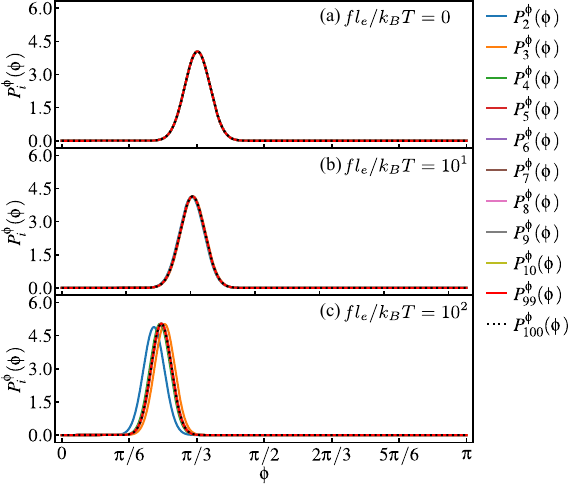}
    \caption{\label{sfig:FRB_Pphii_distribution}Probability distribution of bond angles $P_{i}^\phi(\phi)$ for chains with rigid bonds and deformable bond angles ($\phi_e=60^\circ$, $k_\phi\pi^2/k_BT=10^3$). The chains are subjected to different applied forces: (a) $fl_e/k_BT=0$, (b) $fl_e/k_BT=10^1$, and (c) $fl_e/k_BT=10^2$.}
\end{figure}

Similar to the bond length probability distribution in the fixed bond angle case, the bond angle probability distribution $P_{i}^\phi(\phi)$ converges to a stable distribution $P_\infty^\phi(\phi)$ as the bond index $i$ increases (Fig. \ref{sfig:FRB_Pphii_distribution}). Higher applied forces require a larger bond index for convergence, but the process remains rapid. This converged distribution $P_{i}^\phi(\phi)$ is used to represent the bond angle probability distribution in the chain, and we examine its mean and variance under varying parameters. The average bond angle is
\begin{equation}
    \langle \phi \rangle = \int \phi P_\infty^\phi(\phi)\,d\phi
\end{equation}
and the corresponding variance is
\begin{equation}
    \sigma_\phi^2 = \int (\phi - \langle \phi \rangle)^2P_\infty^\phi(\phi)\,d\phi
\end{equation}

\begin{figure}
    \includegraphics{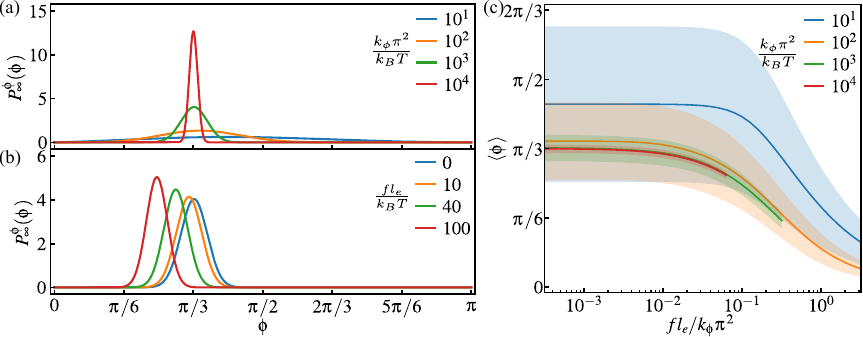}
    \caption{\label{sfig:BondAngleDistribution}Converged bond angle probability distribution in a chain with rigid bonds and deformable bond angles ($\phi_e=60^\circ$). (a) Probability distributions at zero applied force with different bond bending stiffness: $k_\phi\pi^2/k_BT=10^1$, $10^2$, $10^3$, and $10^4$. (b) Probability distributions with bond bending stiffness of $k_\phi\pi^2/k_BT=10^3$, subjected to different forces: $fl_e/k_BT=0$, $10$, $40$, and $100$. (c) Average bond angle and standard deviation as a function of $fl_e/k_\phi\pi^2$.} 
\end{figure}

The converged bond angle distribution $P_\infty^\phi(\phi)$ depends on the bond bending stiffness $k_\phi$ and the applied force $f$. The stiffness $k_\phi$ controls the spread of the distribution. As $k_\phi$ increases, $P_\infty^\phi(\phi)$ narrows around its mean value, indicating reduced bond angle variability (Fig. \ref{sfig:BondAngleDistribution}a). In the limit $k_\phi \pi^2/k_B T \rightarrow\infty$, the bond angle is fixed at $\phi_e$, leading to $P_\infty^\phi(\phi)=\delta(\phi-\phi_e)$. Meanwhile, the force $f$ primarily affects the average bond angle. Increasing $f$ shifts the peak of $P_\infty^\phi(\phi)$ to the left, resulting in a lower average bond angle (Fig. \ref{sfig:BondAngleDistribution}b). Additionally, higher forces slightly sharpen the distribution, further reducing bond angle variability. Fig. \ref{sfig:BondAngleDistribution}c illustrates the dependence of the average bond angle $\langle \phi\rangle$ and the standard deviation $\sigma_\phi$ on the dimensionless parameter $fl_e/k_\phi\pi^2$. As $fl_e/k_\phi\pi^2$ increases, both $\langle \phi\rangle$ and $\sigma_\phi$ decrease, highlighting the combined effect of $f$ and $k_\phi$ on the bond angle behavior.

\begin{figure}
    \includegraphics{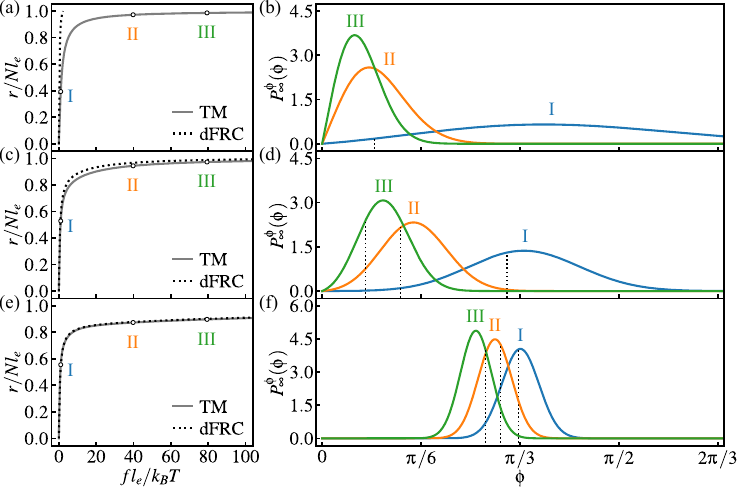}
    \caption{\label{sfig:FRB_ComparisonStatisticalAndSimplified}Comparison of statistical (TM) and dFRC models for chains with rigid bonds and deformable bond angles ($\phi_e=60^\circ$). (a), (c), and (e): Force-extension curves for the bond bending stiffness $k_\phi\pi^2/k_BT=10^1$, $10^2$, and $10^3$, respectively. (b), (d), and (f): Corresponding converged bond angle probability distributions in the statistical model at different applied forces, where the black dotted lines represent the corresponding optimized bond angles in the dFRC model.} 
\end{figure}

Next, we compare how well the dFRC model captures bond angle deformation, using TM calculations as reference. When the bond bending stiffness is very small (eg. $k_\phi\pi^2/k_BT=10^1$), the dFRC model deviates from the TM calculation (Fig. \ref{sfig:FRB_ComparisonStatisticalAndSimplified}a). This occurs because the bond angles are broadly distributed, making the non-fluctuating bond angle assumption of the dFRC model inapplicable (Fig. \ref{sfig:FRB_ComparisonStatisticalAndSimplified}b). However, as $k_\phi\pi^2/k_BT$ increases, the bond angle distribution narrows, and the dFRC model increasingly aligns with the TM calculation. At $k_\phi\pi^2/k_BT=10^3$, it accurately reproduces the chain stretching response (Fig. \ref{sfig:FRB_ComparisonStatisticalAndSimplified}e), although at high applied forces the optimized bond angle in the dFRC model is slightly larger than the most probable value in the distribution (Fig. \ref{sfig:FRB_ComparisonStatisticalAndSimplified}f).  

\section{S7 Applying extensible FJC and FRC models to Carbon chains}
As introduced in Sec. S6.1, the eFJC model proposed by Mao et al. has been widely adopted in polymer modeling. While we have validated its accuracy against TM calculations for freely jointed extensible bonds, its applicability to real polymer chains remains uncertain. Here, we apply the eFJC model to carbon chains, representing them as a series of extensible Kuhn segments. To distinguish this case from the original freely jointed extensible bonds, we refer to it as the eFJC* model. In the eFJC* model, the equilibrium Kuhn length is given as $l_k=\left<r^2\right>/R_0={2l_e\cos{\left(\phi_e/2\right)}}/(1-\cos{\phi_e})$, where $R_0=Nl_e\cos{(\phi_e/2)}$ is the contour length and $\left<r^2\right>=Nl_e^2(1+\cos{\phi_e})/(1-\cos{\phi_e})$ is the mean-square end-to-end distance of the carbon chain at zero force. The stretching stiffness of the Kuhn segment is assumed to match that of carbon-carbon bond \cite{mao2017rupture}. 

For comparison, we also apply the dFRC model to carbon chains, where bonds are extensible but bond angles remain fixed at $\phi_e$. To distinguish this case from the fully deformable dFRC model in Fig. 4a, we refer to it as the extensible Freely Rotating Chain (eFRC) model.
\begin{figure}
    \includegraphics{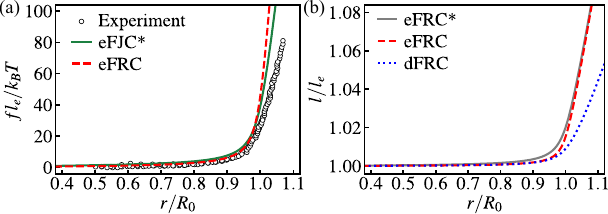}
    \caption{\label{sfig:CarbonChains_sm}Application of the eFJC* and eFRC models to carbon chains. (a) Force-extension curves predicted by FJC* and eFJC* models, compared with experimental data from \cite{wang2013inherent}. (b) Evolution of the Kuhn length $l_k$ in the eFJC* model and the bond length $l$ in the eFRC and dFRC models. $R_0$ is the contour length at zero force.} 
\end{figure}

We compare the predictions of eFJC* and eFRC models with experimental data \cite{wang2013inherent} in Fig. \ref{sfig:CarbonChains_sm}a. While both models improve upon FJC* and FRC by incorporating bond stretching, they still fail to accurately capture the force-extension response for $r/R_0>1.0$ as they neglect bond angle opening. Additionally, we compare the evolution of the Kuhn length in the eFJC* model and the bond length in the eFRC model. As shown in Fig. \ref{sfig:CarbonChains_sm}b, the eFRC* model predicts Kuhn segment extension for $r/R_0>0.8$, and the eFRC model predicts bond extension for $r/R_0>0.9$, aligning with the force regime where FJC* and FRC models break down. However, the predicted Kuhn segment and bond length extensions in these models are much larger than that in the dFRC model, where accounts for both bond extension and bond angle opening. This overestimation leads to inaccurate bond force predictions, ultimately resulting in an incorrect estimation of chain scission. 

\bibliography{bibfile_sm}